\documentclass[preprint2]{aastex}

\def\lae{\lower 2pt \hbox{$\, \buildrel {\scriptstyle <}\over {\scriptstyle \sim}\,$}}

\begin{document}

\title{A multi--Lorentzian timing study of the atoll sources 4U 0614+09 and 4U 1728--34} 

\author{Steve van Straaten\altaffilmark{1}, Michiel van der Klis\altaffilmark{1}, 
Tiziana di Salvo\altaffilmark{1}, Tomaso Belloni\altaffilmark{2}, 
Dimitrios Psaltis\altaffilmark{3}}

\email{straaten@astro.uva.nl}
\altaffiltext{1}{Astronomical Institute, ``Anton Pannekoek'',
University of Amsterdam, and Center for High Energy Astrophysics, 
Kruislaan 403, 1098 SJ Amsterdam, The Netherlands.}

\altaffiltext{2}{Osservatorio Astronomico di Brera, 
Via E. Bianchi 46, I-23807 Merate (LC), Italy.}

\altaffiltext{3}{Center For Space Research,
       Massachusetts Institute of Technology, Cambridge, MA 02139, USA}

\begin{abstract}

We present the results of a multi--Lorentzian fit to the power spectra of 
two kilohertz QPO sources; 4U 0614+09 and 4U 1728--34. This work was 
triggered by recent results of a similar fit to the black--hole 
candidates (BHCs) GX 339--4 and Cyg X--1 by Nowak in 2000. We find that 
one to six Lorentzians are needed to fit the power spectra 
of our two sources. The use of exactly the same 
fit function reveals that the timing behaviour of 4U 0614+09 and 4U 1728--34 
is almost identical at luminosities which are about a factor 5 different.
As the characteristic frequency of the Lorentzians we use the frequency, $\nu_{\rm max}$, 
at which each component contributes most of its variance per log frequency as proposed 
by Belloni, Psaltis \& van der Klis in 2001. When using $\nu_{\rm max}$ instead of the 
centroid frequency of the Lorentzian, the recently discovered hectohertz Lorentzian is 
practically constant in frequency.
We use our results to test the suggestions by, respectively, Psaltis Belloni and 
van der Klis in 1999 and Nowak in 2000 that the two Lorentzians describing the 
high--frequency end of the broad-band noise in BHCs in the low state can be 
identified with the kilohertz QPOs in the neutron star low mass X--ray binaries. The
prediction for the neutron star sources is that if the two kilohertz QPOs are present, 
then these two high--frequency Lorentzians should be absent 
from the broad--band noise. We find, that when the two
kilohertz QPOs are clearly present, the low--frequency part of the power
spectrum is too complicated to draw immediate conclusions from the
nature of the components detected in any one power spectrum. However, the relations 
we observe between the characteristic frequencies of the kilohertz QPOs and the 
band--limited noise, when compared to the corresponding relations in BHCs, hint 
towards the identification of the second--highest frequency Lorentzian in the BHCs 
with the lower kilohertz QPO. They do not confirm the identification of the 
highest--frequency Lorentzian with the upper kilohertz QPO.

\end{abstract}


\section{Introduction}

Low mass X-ray binaries (LMXBs) can be divided into black--hole candidates 
(BHCs) and neutron--star LMXBs. The accretion process in these LMXBs can be studied 
through the timing properties of the associated X-ray emission (see for 
an overview van der Klis 2000). The Fourier power 
spectra of the neutron--star LMXBs contain several timing features; a power--law red 
noise component in the lowest frequency range of the spectrum  ($\nu <$ 1 Hz) called very low 
frequency noise (VLFN), a band--limited noise component (BLN) which is in most cases 
flat at low frequency and steepens to an approximate power--law with an index of about 1 
at higher frequency and several quasi--periodic oscillations (QPOs) at low ($\nu \lae 100$ Hz) and 
high frequencies. The high frequency QPOs, at frequencies from a few hundred Hz to 
more than 1000 Hz are called kilohertz QPOs. The correlation between the timing features at low 
frequency and the spectral properties led to a precise classification of the 
neutron star systems as Z or atoll sources (Hasinger \& van der Klis 1989). 
Note, that the component that we call BLN has been previously called ``high--frequency noise'' 
in atoll sources; this is presumably the same phenomenon that is called ``low--frequency noise'' 
in Z sources (cf. van der Klis 1995).
The timing features of the BHCs include among others a band--limited noise 
component, a bump between 1 and 10 Hz and several QPOs. For the low--frequency ($\nu 
\lae 100$ Hz) part of the power spectrum links between BHCs in the low state and 
neutron--star LMXBs were suggested (e.g., van der Klis 
1994a,b), in particular concerning the band--limited noise and some of the QPOs. 

The improved sensitivity and the discovery of several new timing features 
with the Rossi X--ray Timing Explorer (RXTE) has opened up new 
possibilities for connecting 
the timing properties of the neutron--star LMXBs and the BHCs. In the atoll source 
4U 0614+09 van Straaten et al. (2000, hereafter vS00) found that the relation 
between power density and break frequency of the band--limited noise was 
similar to that established for BHCs in the low state. Wijnands 
\& van der Klis (1999, hereafter WK99) showed that the break frequency of the 
band--limited noise and the centroid frequency of a low--frequency QPO 
follow the same correlation for both the neutron--star LMXBs and the BHCs. 
Psaltis, Belloni \& van der Klis (1999, hereafter PBK99) made a systematic 
study of QPOs and broad--band noise components of the neutron--star LMXBs and 
the BHCs. They suggested the identification of two features by plotting their 
frequencies versus each other. The BHCs' high--frequency bump between 1 and 10 
Hz was tentatively identified with the lower kilohertz QPO in the Z and 
atoll sources and the BHCs' low--frequency QPO with the horizontal branch 
oscillations/low--frequency QPO in the Z/atoll sources. 

Recently, Nowak (2000) described the power spectra of the BHCs GX 339--4 and 
Cyg X--1 with a fit function consisting of four Lorentzian components and 
suggested the identification of the two highest frequency Lorentzians, which 
together describe the high--frequency end of the broad--band noise, with the 
two kilohertz QPOs. The Lorentzian that is used to fit the low--frequency end 
of the broad--band noise has a centroid frequency fixed to zero. The additional
Lorentzian fits a bump at low frequencies.
In the picture of Nowak (2000) for the BHCs the first Lorentzian corresponds to the 
broken power law of WK99, the second Lorentzian to the low--frequency QPO of WK99 and PBK99 
and the third Lorentzian to the high--frequency bump between 1 and 10 Hz of PBK99. 
This work forms 
the original motivation for the current paper. Also triggered by Nowak (2000), 
Belloni, Psaltis \& van der Klis (2001, hereafter BPK01) describe 
a paradigm for fitting the power spectra of low--luminosity bursters and BHCs 
in their low/hard state. They describe the power spectra of these sources with 
a fit function consisting of four Lorentzian components of which in practice 
three are zero--centered. The suggestion of Nowak (2000) might be tested by 
comparing his black hole results with neutron star sources, where both kilohertz QPOs can be clearly 
identified and strong broad--band noise is simultaneously present: the prediction for the neutron star sources
is that if the two kilohertz QPOs are present the two 
high--frequency Lorentzians should be absent from the broad--band noise. In this 
work we test the multi--Lorentzian fit function of Nowak (2000) and BPK01 on two 
of the atoll sources, namely 4U 1728--34 and 4U 0614+09. The timing properties 
at low and high frequencies of 4U 1728--34 and 4U 0614+09 have previously been  
studied in connection with color diagrams by vS00 for 4U 0614+09 
and Di Salvo et al. (2001, hereafter DS01). For both sources strong correlations 
were found between the frequencies of several components and the position of the
source in the color diagram.

We find that instead of four Lorentzians one to six Lorentzians are 
needed to fit the power spectra of 4U 1728--34 and 4U 0614+09. We describe 
these different Lorentzian components in \S 3. In \S 4 we compare our results 
with those of Nowak (2000) and BPK01.

\section{Observations and Data Analysis}

In this work we analyse data from RXTE's proportional--counter array (PCA; for 
instrument information see Zhang et al. 1993) of the neutron--star LMXBs 
4U 1728--34 and 4U 0614+09. For 4U 
1728--34 we used the Fourier power spectra 1 to 19 obtained by and described in 
DS01. The power spectra were constructed by dividing the PCA light curve into 
segments of 256 s and then binning the data in time before Fourier transforming 
such that the Nyquist frequency is always 2048 Hz; the 
normalization of Leahy et al. (1983) was used. Power spectra were combined based 
on the position in the color diagram, which is thought to be an indicator 
of mass accretion rate. The resulting power spectra were then converted 
to squared fractional rms. The Poisson noise estimated between 1200 and 2048 
Hz (where neither noise nor QPOs are known to be present) was subtracted 
before fitting the power spectra. 

For 4U 0614+09 we use a large data set previously studied by vS00 plus
recent observations performed between 2000 September 1 and 4 and 2001 May 24 
and 28. In the September 2000 observations the source was at levels corresponding to 
the lowest mass accretion rate (as inferred from colors and power--spectral properties) 
observed by vS00. In the May 2001 observations the inferred mass accretion rate
was at an even lower level.
Starting May 12, 2000, the propane layer on PCU0, which functions 
as an anti--coincidence shield for charged particles, was lost. This leads to a 
contamination of the data from electrons trapped in the Earth's magnetosphere 
or from solar flare activity. For the September 2000 and May 2001 observations we therefore 
exclude all data with ELECTRON2 $>$ 0.09, where ELECTRON2 is the measured coincidence of events 
between the PCU propane layer and either of the two anodes in the first layer of PCU2, 
the only detector that was switched on in all the observations.  The 
ELECTRON2 screening led to a loss of about 12\% of the data. The loss of the
propane layer in PCU0 also leads to an increased background rate for PCU0. We 
approximately take this into account by adding 30\% of the PCU0 background rate to the 
total background rate. For the colors used above we have excluded PCU0.

In vS00 the 4U 0614+09 data were split into near--continuous time intervals of 
approximately 2500 s which were called observations. For the present work, 
to improve statistics, we constructed representative intervals by adding up several 
observations that showed very similar power spectra in vS00. As all 
characteristic frequencies in the power spectra are correlated with each 
other and with the position of the source in the hardness--intensity diagram 
(parametrized in vS00 by a variable $S_{\rm a}$) we can select the data on 
one of these frequencies. The only power spectral feature that is present in 
all observations is the high--frequency noise. However, at a 
$\nu_{\rm break}$ (the break frequency of the band--limited noise, see for a 
definition vS00) of about 25 Hz and higher the correlation with the other 
frequencies and with the position in hardness--intensity diagram breaks down 
(see figures 3 and 6 in vS00). Therefore, for a $\nu_{\rm break}$ well below 25 
Hz we used $\nu_{\rm break}$ to select the data and for a $\nu_{\rm break}$ of 
about 25 Hz and higher we used the centroid frequencies of the kilohertz QPOs. 
To represent the highest mass accretion rate in 4U 0614+09 (where the 
kilohertz QPOs are mostly absent) we add up the two observations at 
$S_{\rm a}$ = 2.59 and $S_{\rm a}$ = 2.64 in vS00. As in the case of 4U 
1728--34, we number the intervals (1--9) in order of an 
increasing inferred mass accretion rate. We note that we could have selected 
the data based on $S_{\rm a}$ value as was done for 4U 1728--34 (see 
above), but because of the scatter in the correlation of $S_{\rm a}$ with the 
frequencies of the power spectral features (see figures 3 and 4 in vS00), in 4U 
0614+09 this leads to an artificial broadening of the power spectral features 
and in the case of the kilohertz QPOs sometimes even to double peaks.

For each interval of 4U 0614+09 we constructed a power spectrum in the same 
fashion as for 4U 1728--34 (see above), but with a Nyquist 
frequency of 4096 Hz. An interval contains between 53 and 382 power spectra. We 
subtracted the Poisson noise estimated between 2000 and 4000 Hz (as, different 
from 4U 1728--34, QPOs are known to be present between 1200 and 1400 Hz) before 
fitting the power spectra. 

We fitted the power spectra with a sum of Lorentzian components. 
One to six Lorentzian components were needed for a good fit. We only include those 
Lorentzians in the fit whose significance based on the error in the 
power integrated from 0 to $\infty$ is above 3.0 $\sigma$. In addition to the Lorentzians a 
power--law component is used to fit the so called very low--frequency noise 
(VLFN). For the intervals where the kilohertz QPOs have sufficiently high 
frequencies not to interfere with the low--frequency features and vice versa, 
we fit the kilohertz QPOs between 500 and 2048 Hz and then fix the kilohertz 
QPO parameters when we fit the whole power spectra, similar to what was 
done in DS01. This is for computational reasons only; the results are the same 
as those obtained with all parameters free.

We plot the power spectra and the fit functions in the power times frequency 
representation (e.g. Belloni et al. 1997, Nowak 2000), where the power spectral 
density is multiplied with its Fourier frequency. For a fit function consisting of 
many Lorentzians this representation helps to visualize a characteristic frequency 
corresponding to each Lorentzian component (BPK01) namely, the frequency where 
each component contributes most of its variance per logarithmic frequency interval.
This characteristic frequency, is not equal to the centroid frequency, $\nu_0$, of 
the Lorentzian but to $\nu_{\rm max} = \sqrt[]{\nu_0^2 + \Delta^2}$, where $\Delta$ 
is the HWHM of the Lorentzian.
 In this work we therefore use $\nu_{\rm max}$ as characteristic 
frequency for broad features. For narrow features the characteristic frequency 
is nearly the same as $\nu_0$. 
All power spectral features that we find in 4U 0614+09 and 4U 
1728--34 become broader towards lower interval number (see \S 3). 
Several features can be classified as QPOs at higher characteristic frequencies  
but evolve into broad bumps as their frequencies become lower.
Therefore, to be consistent we use $\nu_{\rm max}$ for 
all features. The actual Lorentzian functions fitted were of the form:
$$
P(\nu;\nu_{\rm max},Q,r) = \frac{r^2 \Delta}{(\frac{\pi}{2} + \arctan{2 Q})
(\Delta^2 + (\nu - 2 \Delta Q)^2)}
$$
where $\Delta = \nu_{\rm max}/\sqrt[]{1 + 4 Q^2}$, and $r$ (the fractional rms 
integrated from 0 to $\infty$), $Q$ (the quality factor, defined as $\nu_{\rm 0}/2\Delta$) 
and $\nu_{\rm max}$ were the independent fit parameters.

\section{Results}

The two sources 4U 1728--34 and 4U 0614+09 yielded remarkably similar results. One to six 
Lorentzian components were needed for a good fit. In addition to these 
Lorentzians, intervals 14--19 of 4U 1728--34 and 8 and 9 of 4U 0614+09
needed a power--law component to fit the very low--frequency noise 
(VLFN). We show power spectra and fit functions 
in Figure \ref{fig.powspec_1728} for 4U 1728--34 and in Figure 
\ref{fig.powspec_0614} for 4U 0614+09. The values of $\nu_{\rm max}$ for all 
Lorentzian components are listed in Table \ref{tbl.numax}, the values of $Q$
 are listed in Table \ref{tbl.qvalues} and the values of the integrated fractional 
rms (over the full PCA energy band) are listed in Table \ref{tbl.rms}. The quoted 
errors in $\nu_{\rm max}$, $Q$ and $r$ use $\Delta\chi^{2}$ = 1.0. 

If we compare the $\chi^2$/dof values of our fits with the $\chi^2$/dof values of 
a broken power--law fit (for a description see DS01) to the same intervals there 
seems to be a clear statistical 
preference for the multi--Lorentzian function. For rather similar numbers of degrees 
of freedom (between 131 and 145, and usually higher in the multi--Lorentzian than in 
the broken power--law fit) the $\chi^2$ values of the multi--Lorentzian fits are more 
than 10 lower than of the broken power--law fits in 15 out of 28 cases, whereas the inverse 
is true in only 4 of the 28 cases. The fits are generally better for 
4U 0614+09 (all with a $\chi^2$/dof below 1.9) than for 4U 1728--34 ($\chi^2$/dof below 4.1); 
this could be due to the different methods used to select the data (see \S 2). 

Allmost all fitted power--spectral features (described below) were 
already identified in DS01 and vS00. In DS01 and vS00 the BLN component is described 
with a broken power law, whereas we use a zero-centered Lorentzian. As in DS01 and vS00 
we use Lorentzians to describe two low--frequency QPOs (when present), a broad 
high--frequency feature ($\sim 100$ Hz) and the two kilohertz QPOs. We do not describe 
the behaviour of the VLFN component in this work; for a discussion of this component in 
4U 1728--34 and 4U 0614+09 see respectively DS01 and vS00. In the current work, it only 
affects the power spectra below 0.1--1 Hz.

To confirm the 
identification of the components, we have plotted the $\nu_{\rm max}$ 
of the Lorentzians versus the $\nu_{\rm max}$ of the Lorentzian 
identified as the upper kilohertz QPO ($\nu_{\rm upper kHz}$) in Figure \ref{fig.all_vs_upkilo}. 
Intervals 18 and 19 of 4U 1728--34 are not included in this plot as these intervals only 
showed one Lorentzian. The grey 
symbols mark the 4U 1728--34 points, the black symbols the 4U 0614+09 points. 
In Figure \ref{fig.all_vs_upkilo} five correlations are present, of which the top three 
can be unambiguously identified with the Lorentzians described in \S 3.2--3.4. 
For the other two correlations the identification is more complicated; see \S 3.1.
The points from interval 1 of 4U 0614+09 (from recent observations in 
May 2001) are circled. The Lorentzian at 233 Hz of this interval can be 
identified based on frequency, $Q$ value or fractional rms as either the upper kilohertz QPO 
or the hectohertz Lorentzian. In Tables \ref{tbl.numax}, \ref{tbl.qvalues} and \ref{tbl.rms} 
we list this Lorentzian as the upper kilohertz QPO. In Figure \ref{fig.all_vs_upkilo} and in all
further figures we will use the parameters of this Lorentzian both for the upper kilohertz QPO 
and for the hectohertz Lorentzian. These points will be circled.

The two sources show a remarkable similarity in their behaviour.
This similarity is also shown in the behaviour of the $Q$ values (Table \ref{tbl.qvalues}) 
and the fractional rms's of the
different components (Table \ref{tbl.rms}), although the fractional rms in 4U 0614+09 
is generally higher by about a factor of 1.3 to 2 than that in 4U 1728--34. 
Note that the hydrogen column density, $N_{\rm H}$, is about a factor 10 higher 
for 4U 1728--34 (e.g. Schultz 1999). This leads to the absorbtion of more low--energy 
photons for 4U 1728--34 than for 4U 0614+09 and could increase the factor as most power 
spectral features are stronger at higher energies.
In Figures \ref{fig.rms_vs_upkilo} and \ref{fig.q_vs_upkilo} we plot the fractional rms 
and the $Q$ value of the hectohertz Lorentzian, the lower and upper kilohertz QPOs 
(see below for a description of these components) versus $\nu_{\rm upper kHz}$.

\subsection{The band--limited noise}

The zero--centered Lorentzian which is used to fit the band--limited noise (BLN), 
is present in most of the intervals (except interval 9 of 4U 0614+09). As noted in \S 1, previously 
the BLN in the atoll sources was referred to as HFN (high frequency noise) in the literature, 
but as it has the lowest characteristic frequency of all components that name seems no 
longer appropriate. The characteristic frequency of this component ($\nu_{\rm BLN}$), 
increases with interval number. This increase is halted at interval 13 for 4U 1728--34 
and interval 5 for 4U 0614+09. Here an additional, non zero--centered Lorentzian component 
had to be included in the fit to the BLN. As previously noted by DS01, the characteristic 
frequency of this new component, $\nu_{\rm VLF}$, continues to follow the 
relations between the $\nu_{\rm BLN}$ and the characteristic frequencies of 
the other components (see Fig. \ref{fig.all_vs_upkilo}) as well as the 
relation with interval number. We shall refer to this Lorentzian as the very low--frequency Lorentzian. 
The very low--frequency Lorentzian is generally broader at lower frequencies (see Tables \ref{tbl.numax} and 
\ref{tbl.qvalues}) except for interval 18 and 19 of 4U 1728--34 and 9 of 4U 0614+09 
for which the identification as very low--frequency Lorentzian is ambiguous (see below). 
Note that in the broken power law description, the very low--frequency Lorentzian
is usually not statistically required in 4U 0614+09, whereas it is in 4U 1728--34 (DS01).

In Figure \ref{fig.upkilo_vs_bln} we plot the fractional rms of the BLN versus 
$\nu_{\rm upper kHz}$. In the top panel we plot the fractional rms of the zero--centered 
Lorentzian. In the middle panel we plot the fractional rms of the very low--frequency Lorentzian when 
it is present, and the fractional rms of the zero--centered Lorentzian when it is not. In 
the bottom panel finally we plot the fractional rms of the zero--centered 
Lorentzian and the very low--frequency Lorentzian summed together. In the top two panels the 
relation tends to jump when the very low--frequency Lorentzian appears, where this jump does not occur in
the bottom plot. The relation is also smoother in the bottom plot. This behaviour 
together with the frequency behaviour of the zero--centered Lorentzian and the very low--frequency Lorentzian 
described above could hint at the zero--centered Lorentzian and the very low--frequency Lorentzian being 
two components that together fit one feature. It could however also be that the very low--frequency Lorentzian 
is present in the intervals where we do not detect it and biases our fractional rms and 
$\nu_{\rm max}$ measurements for the zero--centered Lorentzian. This would then explain 
the fractional rms and $\nu_{\rm max}$ connections between the zero--centered Lorentzian 
and the very low--frequency Lorentzian.

We also investigated the possibility that the very low--frequency Lorentzian is a sub--harmonic of the 
low--frequency Lorentzian. We found that in $\nu_{\rm max}$ the deviations
from a harmonic relation are 0.5--4.4 $\sigma$. However, note that in the 
$\nu_{\rm max}$ representation a harmonic relation has no real physical meaning.
In the $\nu_{\rm 0}$ representation, where a harmonic relation has physical meaning, 
the deviations from a harmonic relation between the very low--frequency Lorentzian and the low--frequency 
Lorentzian are 1.9--6.5 $\sigma$.
When we require an harmonic relation between the centroid frequencies of the very low--frequency Lorentzian and the low--frequency Lorentzian in the fits to the power spectra where both features are simultaneously present (5 and 6 for 4U 0614+09 and 9--12 for 4U 1728--34) we find that this significantly worsens the fits (F--test probabilities from 0.05 to $4 \times 10^{-6}$).
Note that for intervals 1, 2, 4, 5 and 6 of 4U 1728--34 there appears to be some residual power 
near the frequency where the sub--harmonic of the low--frequency QPO should be.

In intervals 18 and 19 of 4U 1728--34 and 9 of 4U 0614+09 there is only one Lorentzian present.
Based on the relations of characteristic frequency, $Q$ and rms fractional amplitude with interval 
number, this Lorentzian can be identified as either the zero--centered Lorentzian or the very low--frequency Lorentzian. 
As fitting this 
Lorentzian with a zero--centered Lorentzian provides bad fits, we have listed it as the very low--frequency Lorentzian in 
Tables \ref{tbl.numax} and \ref{tbl.qvalues}, but this identification is uncertain. 
Contrary to all other components, the characteristic frequency of this component decreases with interval 
number (see Tables \ref{tbl.numax} and \ref{tbl.qvalues}).

\subsection{The low--frequency Lorentzian}

The second Lorentzian at low frequencies (which we shall refer to as the
low--frequency Lorentzian) was present in intervals 1--12 for 4U 1728--34 and 1--6 
for 4U 0614+09 (see Figures \ref{fig.powspec_1728} and \ref{fig.powspec_0614}).
The characteristic frequency of this component, $\nu_{\rm LF}$, also increases 
with interval number (see Table 
\ref{tbl.numax}). In 4U 1728--34 $\nu_{\rm LF}$ ranges from 11.5--46.7 Hz, 
in 4U 0614+09 from 2.0--44.6 Hz. The low--frequency Lorentzian is too broad to 
be classified as a QPO ($Q < 2$) at low frequencies ($\nu_{\rm LF} <$ 25 Hz; 
see Tables \ref{tbl.numax} and \ref{tbl.qvalues}) and $Q$ only exceeds 2
at higher frequencies. Note that the low--frequency Lorentzian might 
be present in interval 6 of 4U 0614+09 ($\nu_{\rm LF}$ = 60.5$^{+1.6}_{-2.2}$ 
Hz); as the significance based on the error in the integrated power is only 
2.4 $\sigma$ we have not included this Lorentzian in the fit. 

\subsection{The hectohertz Lorentzian}

The fourth component, which we shall refer to as the hectohertz Lorentzian, 
is fitted with a Lorentzian with a characteristic frequency of the order of 
100 Hz. The hectohertz Lorentzian is only absent in intervals 18 and 19 of 
4U 1728--34, at the highest inferred mass accretion rate. The Lorentzian at 233 Hz
in interval 1 of 4U 0614+09, at the lowest inferred mass accretion rate, can be
identified as either the hectohertz Lorentzian or the upper kilohertz QPO. 
The hectohertz Lorentzian 
is broad at low interval numbers and becomes narrower at high interval numbers (see 
Table \ref{tbl.qvalues}). This is a good example of how the characteristic frequency 
$\nu_{\rm max}$ of a Lorentzian is determined by $\nu_0$ and $\Delta$. 
In Figure \ref{fig.hecto} we plot $\nu_{\rm max}$, 
$\nu_{\rm 0}$ and $\Delta$ versus the centroid frequency of the upper 
kilohertz QPO. As $\nu_0$ increases, $\Delta$ falls such that $\nu_{\rm max}$ is nearly 
constant.

\subsection{The kilohertz QPOs}

The fifth and sixth Lorentzians were used to fit the pair of kilohertz QPOs. 
This pair is present in intervals 10--17 for 4U 1728--34 and 6--8 for 4U 
0614+09. In interval 1 of 4U 0614+09 there is a Lorentzian present at $\sim 20$
Hz. Based on its frequency this Lorentzian can be tentatively identified with the relation 
which PBK99 associate with the lower kilohertz QPO. However, this Lorentzian is much broader 
and stronger than any of the ``normal'' lower kilohertz QPOs in 4U 0614+09 and 4U 1728--34 
(see Figures \ref{fig.rms_vs_upkilo} and \ref{fig.q_vs_upkilo}); its broad shape is similar 
to the corresponding feature seen at similar frequencies in the low--luminosity
bursters (BPK01). In the other intervals (except 18 and 19 for 4U 1728--34) there is 
only one kilohertz QPO present, which can be identified as the upper kilohertz 
QPO based on the correlations of its frequency with the frequency of the other 
features (see Fig. \ref{fig.all_vs_upkilo}) or interval number (see DS01).
The Lorentzian at 233 Hz in interval 1 of 4U 0614+09 can be
identified as either the hectohertz Lorentzian or the upper kilohertz QPO.
Both kilohertz QPOs are broader at lower frequencies (see Tables \ref{tbl.numax} 
and \ref{tbl.qvalues}). At frequencies below $\sim 550$ Hz the upper kilohertz QPO 
becomes too broad to be classified as a real QPO ($Q < 2$).

\section{Comparison with other sources}
 
A way to compare the results of our multi--Lorentzian fit to the two atoll 
sources with previous fits of this function to other neutron--star LMXBs and BHCs 
(Shirey 1998, Nowak 2000, BPK01) is to plot the characteristic frequencies of 
the different Lorentzians components versus the frequency of one Lorentzian 
component which has already been identified for all these sources. There are  
three obvious candidates to plot everything against that have already been identified 
by WK99 and PBK99 (see the introduction): the band--limited noise component, the low--frequency 
Lorentzian and the lower kilohertz QPO. Here we choose to present the results of 
comparing to the band--limited noise component. If we use the characteristic 
frequency of the low--frequency Lorentzian instead, our conclusions do not change. 
If we use the characteristic frequency of the lower kilohertz QPO our points fall 
on the relations shown in BPK01; any differences 
between our conclusions and those of BPK01 are discussed below. 

The large advantage of using the band--limited noise component is that in our 
data this component is present in all spectra, whereas the lower kilohertz 
QPO is only present in 11 (perhaps 12, see \S3.4) and the low--frequency Lorentzian 
in 18 out of the 28 power spectra. Also in the other sources the band--limited noise 
component can almost always be clearly identified, where for the other peaks there can be 
ambiguity about which peak is which (see, e.g., Homan et al. 2001). A disadvantage of 
using the band--limited noise component is that in both 4U 1728--34 and 4U 0614+09 
at a certain point the frequency of the very low--frequency Lorentzian ``takes over'' from the frequency of the 
zero--centered Lorentzian 
(see \S 3.1). So, just plotting the characteristic frequencies of the different 
Lorentzian components versus $\nu_{\rm BLN}$ leads to a confusing picture 
where the correlations break down, i.e. at the intervals where both the zero--centered 
Lorentzian and the very low--frequency Lorentzian are present. This can be resolved by using 
$\nu_{\rm VLF}$ as the characteristic frequency of the the band--limited noise if 
the very low--frequency Lorentzian is present, and $\nu_{\rm BLN}$ otherwise. We call the thus defined 
characteristic frequency of the band--limited noise $\nu_{\rm band}$. 
Although the physical connection between the zero--centered Lorentzian and the 
very low--frequency Lorentzian is uncertain (see \S 3.1), the smooth run of $\nu_{\rm BLN}$ and 
$\nu_{\rm VLF}$ with upper kilohertz QPO frequency visible in Fig. \ref{fig.all_vs_upkilo} 
indicates that for 4U 0614+09 and 4U 1728--34 $\nu_{\rm band}$ could serve as horizontal 
coordinate to plot everything against as well as $\nu_{\rm upper kHz}$.
We caution however, that due to the extensive data sets available for our two sources we are able 
to keep track of $\nu_{\rm band}$ through the transformation from zero--centered Lorentzian 
to very low--frequency Lorentzian. With less data, confusion is likely. The correlation of $\nu_{\rm band}$ with 
the other frequencies breaks down at interval 9 of 4U 0614+09 (cf. \S 3.1). However, as this 
is only one very extreme point this is not a real problem for our present purpose. We have 
excluded intervals 18 and 19 of 4U 1728--34 from the correlations, as for these intervals 
only one component, the very low--frequency Lorentzian, is present. But note that in these intervals also there is evidence
of a break down in the correlations with $\nu_{\rm band}$ (\S 3.1).

In Figure \ref{fig.comp} we compare our results with those of Nowak (2000; the BHCs GX 339--4, 
Cyg X--1) and BPK01 (the low--luminosity bursters 1E 1724--3045, GS 1826--24, SLX 1735--269 
and the BHC XTE J1118+480) by plotting all characteristic frequencies 
versus $\nu_{\rm band}$. The points for which we use $\nu_{\rm BLN}$ 
as $\nu_{\rm band}$ are black, those for which we use $\nu_{\rm VLF}$ are grey.
We have also included the results for Cir X--1 from Shirey 
(1998) that were also used in PBK99. That author also used a zero--centered Lorentzian 
to fit the broad--band noise, and fitted two peaked features  with Lorentzians. 
The only difference with Nowak (2000) and BPK01 is that the high--frequency tail 
in Shirey (1998) was fitted with a power law, whereas Nowak (2000) and BPK01 use 
an additional Lorentzian. The grey symbols in Fig. \ref{fig.comp} represent the 
results from Shirey (1998), Nowak (2000) and BPK01, the black symbols our 
results on 4U 1728--34 and 4U 0614+09. For clarity Figure \ref{fig.comp} is 
split up in three panels. In the bottom panel we plot the characteristic 
frequency of the low--frequency Lorentzian ($\nu_{\rm LF}$), versus 
$\nu_{\rm band}$ (the WK99 relation). In this plot we have included the points 
from WK99 without the 4U 1728--34 and 4U 0614+09 points. However, note that 
WK99 use a broken power--law instead of a 
zero--centered Lorentzian to fit the band--limited noise. This will lead to 
systematic deviations in characteristic frequency (see BPK01). Also note that
for both 1E 1724--3045 and GS 1826--24 there are usually two low frequency
peaks present (see BPK01); a narrow peak and a broad component at a slightly higher 
frequency. If both these peaks are present we use the $\nu_{\rm max}$ of the broad 
peak to compare with 4U 1728--34 and 4U 0614+09 based on the strong similarities of the 
broad peak with the low frequency Lorentzian in these sources (see also Figure 
\ref{fig.cygx1_1E1724_4U0614}).

Most of our results on 4U 1728--34 and 4U 0614+09 coincide with the WK99 data. A few 
points lie slightly below the WK99 relation just as the points of 
GX 339--4, Cyg X--1 of Nowak (2000) and 1E 1724--3045, GS 1826--24 and SLX 1735--269 
of BPK01. This may be due to the systematic frequency deviations between
a zero--centered--Lorentzian and a broken power--law (see BPK01). The XTE J1118+480 
points of BPK01 and the Cir X--1 points of Shirey (1998) are well below the WK99
relation but may line up with each other. Note that the QPO in Cir X--1 was previously identified as 
$\nu_{\rm LF}$ with the PBK99 relation (see PBK99).

In the middle and top panels we plot the characteristic frequencies of the 
hectohertz Lorentzian and both kilohertz QPOs versus $\nu_{\rm band}$ for 
4U 1728--34 and 4U 0614+09. We have fitted the correlations of the 
characteristic frequencies of the hectohertz Lorentzian and the upper 
kilohertz QPO to $\nu_{\rm band}$ with power--laws in order to extrapolate these 
correlations to lower frequencies (dashed lines in Fig. \ref{fig.comp}).
For the lower kilohertz QPO the range in $\nu_{\rm band}$ 
from these two sources alone is too small to draw strong conclusions from the extrapolation.
For these fits we excluded interval 9 of 4U 0614+09 where the correlations 
between $\nu_{\rm band}$ and the frequencies of the other Lorentzians break 
down (see Fig. \ref{fig.all_vs_upkilo}). We also exclude interval 1 of 4U 0614+09 
from the fit (although these points are shown in the Figure) as for this interval 
the identification of the hectohertz Lorentzian
and the lower and upper kilohertz QPO are uncertain (see \S 3).
We use these power--law fits only to get an indication for what the characteristic 
frequencies for the hectohertz Lorentzian and the upper kilohertz QPO would be 
expected to be at low frequencies. The 
difference between the two panels is that in the middle panel we compare with the 
characteristic frequency of the Lorentzian ($\nu_\ell$ in BPK01) that PBK99 (for the points 
from Shirey 1998), Nowak (2000) and BPK01 identify with the lower kilohertz QPO and in the 
top panel with the Lorentzian ($\nu_u$ in BPK01) that Nowak (2000) and BPK01 identify with 
the upper kilohertz QPO. We find that also when plotting versus $\nu_{\rm band}$ the relation 
between $\nu_\ell$ and $\nu_{\rm band}$ (middle frame) is similar to the relation 
of the lower kilohertz QPO with $\nu_{\rm band}$. Also the $\sim 20$ Hz Lorentzian of interval 
1 of 4U 0614+09 falls fairly close to the relation between $\nu_\ell$ and $\nu_{\rm band}$. 
However, more work needs to be done to connect both relations.  

Concerning the top panel, Nowak (2000) and BPK01 found that the BHC points (GX 339--4, Cyg 
X--1 and XTE J1118+480) as well as the SLX 1735--269 point followed the extension 
to low frequencies of the ``lower versus upper kilohertz QPO frequency'' of PBK99. The 
points of 1E 1724--3045 and GS 1826--24 were above the relation. We find that the BHC 
points (GX 339--4, Cyg X--1 and XTE J1118+480) as well as the single point from SLX 
1735--269 lie well below the extrapolated relation of $\nu_{\rm band}$ with the 
characteristic frequency of the upper kilohertz QPO; the points might be consistent 
with the extrapolation of the lower kilohertz QPO relation. We can therefore not 
identify the highest frequency Lorentzian in these sources with the upper kilohertz QPO 
based on this relation. Note again that the identification of $\nu_{\rm band}$ might 
be wrong in sources for which only litle data has been analyzed (see above).
The points of GS 1826--24 are well above the relation and only 
the points of 1E 1724--3045 fall in the right range. However, the 1E 1724--3045 points also 
fall in the right range to be identified with the hectohertz Lorentzian. The same conclusion 
can be drawn for the 233 Hz Lorentzian of interval 1 of 4U 0614+09. 

Interval 1 of 4U 0614+09 shows a $\nu_{\rm band}$ similar to that of the Cyg X--1 observation
from Nowak (2000) and to that of observations of the low luminosity bursters 1E 1724--3045 and GS 1826--24 
from BPK01. In Figure \ref{fig.cygx1_1E1724_4U0614} we plot the power spectra of 4U 0614+09, 1E 1724--3045 
and Cyg X--1 that have a $\nu_{\rm band}$ of about 0.3 Hz. For 1E 1724--3045 we use the same data 
that made up interval E of BPK01 and for Cyg X--1 we use the RXTE observations on 1996 October 22, the
same as used by Nowak (2000). We constructed the power spectra in the same fashion as for 4U 
1728--34 and 4U 0614+09 (see \S 2). The power spectra of 4U 0614+09 and 1E 1724--3045 are almost identical 
at both low and high frequencies. The power spectrum of Cyg X--1 is only similar to those of the neutron 
stars at low frequencies. The two high frequency peaks are both weaker relative to the low frequency peaks 
and about a factor of 4 lower in characteristic frequency than those in 4U 0614+09 and 1E 1724--3045. An 
explanation for this might be that the characteristic frequencies of the high frequency peaks in the spectrum 
scale with the mass of the central object and that the low frequency peaks are independent of this mass.

\section{Discussion}

\subsection{Comparison between 4U 1728--34 and 4U 0614+09}

The use of exactly the same fit function for both 4U 1728--34 and 4U 0614+09 
shows the similar timing behaviour of these two atoll sources (see \S 3). 
Especially the frequencies and the $Q$ values are almost identical. This similarity 
is quite remarkable, as the X--ray luminosity is about a factor 5 larger for 4U 
1728--34 (Ford et al. 2000). Note that the luminosities for 4U 1728--34 and 4U 0614+09 
in Ford et al. (2000) are based on distance to these sources calculated from type I 
X--ray bursts. The luminosity is thought to reach the Eddington 
limit in the so called radius expansion bursts (Lewin, van Paradijs \& Taam 1993).
For 4U 1728--34 a radius expansion burst was used (Foster et al. 1986) but for
4U 0614+09 no radius expansion bursts have been observed, therefore only an upper limit 
on the distance was derived by assuming that the flux is less than the Eddington limit 
(Brandt et al. 1992). So the luminosity in Ford et al. (2000) for 4U 0614+09 is also 
an upper limit. The fractional rms in 4U 1728--34 is systematically 
lower than that in 4U 0614+09, which may be related to its higher luminosity. 
Note that the data are inconsistent with a model in which the kilohertz QPO 
amplitudes are the same and the higher luminosity of 4U 1728--34 is caused by 
an additional source of X--rays unrelated to the kilohertz QPOs; then the rms 
values in 4U 1728--34 would be expected to be a factor of 5 lower, while they 
are only lower by about a factor of 1.3 to 2. Another stunning difference between 
4U 1728--34 and 4U 0614+09 is that in the whole RXTE data set on 4U 0614+09 obtained 
until now ($\sim 900$ ks) no type I X--ray bursts have been observed, whereas 4U 
1728--34 is a well known burster that shows type I X--ray bursts in all parts of 
the color--color diagram (van Straaten et al. 2001, Franco 2001). Note that although
no type I X--ray bursts have been observed for 4U 0614+09 with RXTE, two X--ray bursts 
from this source have been previously detected. One with the OSO--8 satellite (Swank et 
al. 1978) and one with the GRANAT satellite (Brandt et al. 1992).

\subsection{The hectohertz Lorentzian}

The hectohertz Lorentzian has been identified in several other neutron star LMXBs (SAX J1808.4-3658, 
4U 1705--44; Wijnands \& van der Klis 1998). Sunyaev \& Revnivtsev (2000) use a sample of 9 BHCs 
and 9 neutron star LMXBs (including SAX J1808.4-3658, 4U 1705--44, 4U 0614+09 and 4U 1728--34) to 
show that neutron star systems show significant broad--band noise above several hundred Hz where BHC 
do not. The hectohertz Lorentzian is one of the contributers to this excess of power in SAX J1808.4-3658, 
4U 1705--44, 4U 0614+09 and 4U 1728--34 and could be in the other sampled neutron star LMXBs.  

The hectohertz Lorentzians could be related to the $\sim 67$ Hz QPO in the BHC GRS 1915+105 (Morgan, 
Remillard \& Greiner 1997) and the $\sim 300$ Hz QPO in the BHC GRO J1655--40 (Remillard et al. 1999). 
These oscillations also have stable frequencies and fall in the same frequency range as the 
hectohertz Lorentzians we observe. However, these QPOs are much narrower ($Q \sim 5$ for the $\sim 
300$ Hz QPO and $Q \sim 20$ for the $\sim 67$ Hz QPO) and have much weaker rms fractional amplitudes 
($\sim$1 \%). Models to explain these stable QPOs in the BHCs all invoke strong gravity near the central 
black hole (see for a review Cui, Chen \& Zhang 2000).

Recently, Fragile, Mathews \& Wilson (2001) made a tentative identification of the 9 Hz 
QPO in the BHC GRO J1655--40 (Remillard et al. 1999) with the orbital frequency at the Bardeen--Petterson 
transition radius (e.g. Bardeen \& Petterson 1975). Based on the similarities between the hectohertz 
Lorentzian in the neutron star LMXBs and the 9 Hz QPO in GRO J1655--40 they also suggest this 
identification for the hectohertz Lorentzian. For 4U 0614+09 and 4U 1728--34 we can confirm that 
in indeed several properties of the hectohertz Lorentzian are quite similar to those of the 9 Hz QPO in GRO 
J1655--40. As in the case of the 9 Hz QPO, the hectohertz Lorentzian always has $Q < 3$ and for both features 
the characteristic frequency is nearly constant whilst the frequencies of other components change.  

\subsection{Comparison with BHCs and low--luminosity bursters}

As noted in \S 1, if the two highest frequency QPOs in the BHCs 
could be identified with the lower and upper kilohertz QPO in the atoll sources (as suggested 
by respectively PBK99 and Nowak 2000), one would predict that in 4U 1728--34 and 4U 0614+09 the 
two high--frequency Lorentzians should be absent from the broad--band noise if the two kilohertz 
QPOs are present. Our results show a more complicated picture; when the two kilohertz 
QPOs are clearly present, then in addition to the zero--centered Lorentzian and the low--frequency
Lorentzian, two more Lorentzian components are present, the very low--frequency Lorentzian and the hectohertz Lorentzian. 
These Lorentzians however have quite a special character. The characteristic 
frequency of the very low--frequency Lorentzian is strongly connected with the characteristic frequency of 
the zero--centered Lorentzian and might be part of the BLN (see \S 3.1), which is not clear 
for the two highest--frequency Lorentzians describing the broad--band noise in the BHCs. The 
hectohertz Lorentzian is almost constant in frequency with a frequency of about 150 Hz. The 
highest--frequency Lorentzian describing the broad--band noise in the BHCs might also be constant in 
frequency at about 30 Hz, but there is currently not enough data available for the BHCs to confirm this. 
The factor of $\sim 5$ difference in frequency might then be caused by a scaling with the mass of the 
compact object which is about a factor of 5 lower for the neutron stars.

The power spectra of the low--luminosity bursters and BHCs studied in Nowak (2000) 
and BPK01 showed four Lorentzian components three of which in practice are zero--centered. 
The atoll sources 4U 1728--34 and 4U 0614+09 also show four 
Lorentzian components at low inferred mass accretion rates (intervals 1--8 
of 4U 1728--34 and 1--4 of 4U 0614+09). So can we identify all four peaks in these 
intervals of 4U 1728--34 and 4U 0614+09 with the four peaks in the BHCs and the 
low--luminosity bursters? The first two Lorentzian components in the low--luminosity 
bursters, BHCs and the atoll sources 4U 1728--34
and 4U 0614+09 may have the same physical origin as they also show a similar 
relation between their respective characteristic frequencies (the WK99 relation, see Fig.
\ref{fig.comp}, bottom panel). Interval 1 of 4U 0614+09 is almost identical to that 
of the low--luminosity bursters. However, because interval 1 of 4U 0614+09 is an
extreme point in the current dataset available for 4U 0614+09, its two highest--frequency Lorentzians 
cannot be linked directly with the two highest frequency Lorentzians of intervals 2--4. More observations
of 4U 0614+09 with a $\nu_{\rm band}$ between 0.3 and 1.4 Hz are needed to connect these points.

The third Lorentzian of the BHCs and low--luminosity 
bursters however, behaves completely differently from the third Lorentzian in 4U 1728--34 
and 4U 0614+09 (except for interval 1). Where the characteristic frequency of the third Lorentzian in 4U 
1728--34 and 4U 0614+09 (i.e. the hectohertz Lorentzian) is almost constant at about 150 Hz, 
the third Lorentzian in the 
BHCs and low--luminosity bursters varies over several orders of magnitude (Fig. 
\ref{fig.comp}, middle panel). Instead, the PBK99 relation as well as the relation 
between the characteristic frequency of the third Lorentzian in the BHCs and 
low--luminosity bursters with $\nu_{\rm band}$  suggest the identification of this Lorentzian 
with the lower kilohertz QPO that is 
present in intervals 10--17 of 4U 1728--34 and intervals 6--8 of 4U 0614+09. 
In the middle panel of Fig. \ref{fig.comp} we have illustrated this with a dotted line that
indicates a power law fit to the characteristic frequencies of the lower kilohertz QPO of 
4U 1728--34/4U 0614+09 and the third Lorentzian of the low--luminosity bursters to $\nu_{\rm band}$.

Based on 
the similarity between the relation between the characteristic frequencies of the third and 
the fourth Lorentzians in the BHCs GX 339--4 and Cyg X--1
and the relation between the characteristic frequencies of the lower and upper kilohertz 
QPOs, Nowak (2000) suggested the identification of the fourth Lorentzian with the upper 
kilohertz QPO. Indeed the upper kilohertz QPO is also the fourth peak in the intervals 
at lower inferred mass accretion rates,
where 4U 1728--34 and 4U 0614+09 mimic the BHCs most. However, the relation between the 
characteristic frequency of the fourth Lorentzian and $\nu_{\rm band}$ does not confirm 
this identification (see \S 4). For the low--luminosity bursters 1E 1724--3045 and GS 
1826--24 studied in BPK01 the identification of the fourth peak with the upper kilohertz 
QPO is even harder (see \S 4). But keep in mind that due to the extensive data sets available 
for our two sources we are able to keep track of $\nu_{\rm band}$ through the transformation 
from zero--centered Lorentzian to very low--frequency Lorentzian. With less data, confusion might occur.

\subsection{Summary}

\begin{itemize} 

\item 	The use of exactly the same fit function for 4U 0614+09 and 4U 1728--34 shows 
      	that these two atoll sources show a remarkably similar timing behaviour at 
	luminosities that differ by a factor of 5.

\item   Interval 1 of 4U 0614+09 shows a remarkably similar power spectrum to that of 
	the low--luminosity bursters. Unfortunately, this is an extreme point in the current 
	dataset available for 4U 0614+09 and therefore the two highest frequency Lorentzians 
	cannot be linked directly with the high frequency Lorentzians of the other intervals. 
	More observations of 4U 0614+09 with a $\nu_{\rm band}$ between 0.3 and 1.4 Hz are needed 
	to complete the picture.

\item   When the two kilohertz QPOs are clearly present, the low--frequency 
	part of the power spectrum is too complicated to draw immediate conclusions from the
	nature of the components detected in any one power spectrum. The relation of characteristic 
	frequency of respectively the two high--frequency 
        Lorentzians with the characteristic frequency of the band--limited noise ($\nu_{\rm band}$) hints 
        towards the identification of the second--highest frequency Lorentzian in the BHCs with the lower 
        kilohertz QPO (as suggested by PBK99) but can not confirm the identification of the highest 
        frequency Lorentzian with the upper kilohertz QPO (as suggested by Nowak 2000).

\item 	When using $\nu_{\rm max}$ instead of the centroid frequency, the recently 
      	discovered hectohertz Lorentzians are almost stable in frequency.

\item   All Lorentzian components in 4U 0614+09 and 4U 1728--34 become broader as their 
	frequency decreases. This might explain why many Lorentzians in other sources are
	zero--centered when present at low frequencies.
	
\item	For 4U 0614+09 the use of the multi--Lorentzian fit function clearly shows the 
	presence of the band--limited noise QPO, where in the broken power law 
	description, the very low--frequency Lorentzian is usually not statistically required. 
	The similarity between the very low--frequency Lorentzian in 4U 0614+09 and 4U 1728--34, 
	where the very low--frequency Lorentzian is statistically required in both descriptions, 
	confirms the identification of the very low--frequency Lorentzian in 4U 0614+09.

\end{itemize}

\section{Acknowledgements}

This work was supported by NWO SPINOZA grant 08--0 to E.P.J. van den Heuvel, 
by the Netherlands Organization for Scientific Research (NWO), and by 
the Netherlands Research School for Astronomy (NOVA). 
This research has made use of data obtained through
the High Energy Astrophysics Science Archive Research Center Online Service, 
provided by the NASA/Goddard Space Flight Center. We would like to thank 
Bob Shirey for providing us with a table of the Cir X--1 data.


\begin{center}
\begin{deluxetable}{lccccccc}
\tabletypesize{\footnotesize}
\tablewidth{0pt}
\tablecaption{Characteristic frequencies of the multi--Lorentzian fit for 
4U 1728--34 \& 4U 0614+09\label{tbl.numax}} 
\tablehead{ 
\colhead{Interval} & \colhead{zc Lorentzian} & 
\colhead{VLF Lorentzian} & \colhead{LF Lorentzian} & 
\colhead{hHz Lorentzian} & 
\colhead{Lower kHz QPO} & \colhead{Upper kHz QPO} \\
\colhead{Number} & \colhead{$\nu_{\rm BLN}$ (Hz)} & 
\colhead{$\nu_{\rm VLF}$ (Hz)} & \colhead{$\nu_{\rm LF}$ (Hz)} & 
\colhead{$\nu_{\rm hHz}$ (Hz)} & \colhead{$\nu_{\rm lower kHz}$ (Hz)} & 
\colhead{$\nu_{\rm upper kHz}$ (Hz)}  
}

\startdata
\multicolumn{7}{c}{4U 1728--34}\\
\hline
1  & 1.976$\pm$0.057       & --	                      & 11.47$\pm$0.16  & 178$\pm$35       & --                    & 399$\pm$13\\
2  & 2.077$\pm$0.046       & --	                      & 11.75$\pm$0.13  & 173$\pm$22       & --                    & 416$\pm$12\\
3  & 2.983$\pm$0.083       & --	                      & 16.19$\pm$0.25  & 203$\pm$28       & --                    & 497$\pm$14\\
4  & 3.976$\pm$0.060       & --	                      & 19.72$\pm$0.17  & 175.0$\pm$5.3    & --                    & 519.1$\pm$5.1\\	
5  & 4.184$\pm$0.059       & --	                      & 20.12$\pm$0.16  & 175.3$\pm$6.1    & --                    & 521.4$\pm$5.2\\	
6  & 4.56$\pm$0.12         & --	                      & 21.56$\pm$0.33  & 198$\pm$13       & --                    & 531.7$\pm$8.0\\	
7  & 12.3$\pm$0.9          & --	               	      & 27.5$\pm$1.0    & 175$\pm$14       & --                    & 705.7$\pm$8.1\\
8  & 16.63$\pm$0.60        & --	                      & 31.29$\pm$0.63  & 152.3$\pm$6.5    & --                    & 732.6$\pm$3.1\\	
9  & 14.2$\pm$4.5          & 16.20$\pm$0.46           & 36.26$\pm$0.47  & 130.7$\pm$5.9    & --                    & 791.1$\pm$2.2\\
10 & 19.5$\pm$2.3          & 19.39$\pm$0.31           & 42.14$\pm$0.77  & 136.9$\pm$4.5    & 513$\pm$18            & 849.5$\pm$2.0\\	
11 & 21.2$\pm$2.2          & 20.96$\pm$0.23           & 45.52$\pm$0.66  & 127.3$\pm$3.2    & 561$\pm$11            & 875.7$\pm$1.6\\	
12 & 18.8$\pm$2.6          & 23.08$\pm$0.29           & 46.70$\pm$0.91  & 129.7$\pm$3.6    & 604$\pm$14            & 907.6$\pm$2.5\\	
13 & 11.6$\pm$1.2          & 30.61$\pm$0.53           & --	        & 130.9$\pm$2.9    & 680$\pm$10            & 951.0$\pm$3.7\\	
14 & 16.9$\pm$1.8          & 40.64$\pm$0.47           & --	        & 147.1$\pm$5.3    & 754.0$\pm$3.7         & 1056.2$\pm$8.2\\	
15 & 13.3$\pm$1.0          & 42.21$\pm$0.29           & --	        & 164.3$\pm$5.3    & 775.3$\pm$1.7         & 1107.5$\pm$7.2\\	
16 & 14.5$\pm$1.5          & 43.39$\pm$0.32           & --	        & 151.6$\pm$7.8    & 819.9$\pm$4.1         & 1134$\pm$12\\	
17 & 12.9$^{+5.7}_{-3.5}$  & 43.21$\pm$0.98           & --	        & 224$\pm$27       & 879.2$\pm$3.0         & 1161$\pm$16\\
18 & --		           & 33.4$\pm$3.6$^{1}$       & --	        & --	           &  --	           &  --\\
19 & --		           & 20.9$\pm$3.5$^{1}$       & --	        & --	           &  --	           &  --\\
\hline
\multicolumn{7}{c}{4U 0614+09}\\
\hline
1  & 0.3054$\pm$0.0046     & --	                      & 2.057$\pm$0.014 &  --	           & 19.97$\pm$0.46$^{1}$  & 233.4$\pm$8.6$^{1}$\\
2  & 1.434$\pm$0.020       & --	                      & 9.558$\pm$0.076 & 157$\pm$12       & --                    & 369.7$\pm$7.5\\
3  & 3.47$\pm$0.14         & --	                      & 16.39$\pm$0.43  & 153$\pm$12       & --                    & 481$\pm$15\\
4  & 10.74$\pm$0.71        & --	                      & 25.38$\pm$0.76  & 153.6$\pm$7.7    & --                    & 629.6$\pm$4.3\\
5  & 7.0$^{+2.6}_{-1.7}$   & 17.54$\pm$0.46           & 37.70$\pm$0.63  & 121.8$\pm$5.3    & --                    & 755.8$\pm$4.1\\
6  & 9.1$\pm$1.7           & 21.11$\pm$0.37           & 44.4$^{+1.2}_{-0.6}$ & 121.0$\pm$3.3    & 517$\pm$12            & 833.2$\pm$4.2\\
7  & 11.4$^{+3.0}_{-2.0}$  & 26.17$\pm$0.42           & --	        & 132.3$\pm$4.7    & 607.9$\pm$1.7         & 925.9$\pm$3.5\\
8  & 21.0$\pm$5.0          & 31.17$\pm$0.41           & --	        & 148.9$\pm$5.0    & 754.8$\pm$2.4         & 1137.1$\pm$7.0\\
9  & --                    & 23.77$\pm$0.92$^{1}$           & --	        & 117.5$\pm$1.5    & --                    & 1273.6$\pm$9.5\\
\enddata

\tablecomments{Listed are the characteristic frequencies ($\equiv 
\nu_{\rm max}$) of the different Lorentzians described in \S 3. The quoted 
errors in $\nu_{\rm max}$ use $\Delta\chi^{2}$ = 1.0.\\
$^{1}$ The identification of this Lorentzian is not clear, see \S 3. 
}

\end{deluxetable}
\end{center}

\begin{center}
\begin{deluxetable}{lcccccc}
\tabletypesize{\footnotesize}
\tablewidth{0pt}
\tablecaption{$Q$ values of the multi--Lorentzian fit for 
4U 1728--34 \& 4U 0614+09\label{tbl.qvalues}} 
\tablehead{ 
\colhead{Interval} & 
\colhead{VLF Lorentzian} & \colhead{LF Lorentzian} & 
\colhead{hHz Lorentzian} & 
\colhead{Lower kHz QPO} & \colhead{Upper kHz QPO} \\
\colhead{Number} &  & & & & }

\startdata
\multicolumn{6}{c}{4U 1728--34}\\
\hline
1  & --	                      & 0.589$\pm$0.037     & 0.03$\pm$0.15     & --                    & 1.66$^{+0.64}_{-0.41}$\\
2  & --	                      & 0.578$\pm$0.030     & 0.09$\pm$0.11     & --                    & 1.46$\pm$0.26\\
3  & --	                      & 0.437$\pm$0.029     & 0.51$\pm$0.14     & --                    & 1.77$\pm$0.42\\
4  & --	                      & 0.594$\pm$0.021     & 0.769$\pm$0.073   & --                    & 1.81$\pm$0.13\\	
5  & --	                      & 0.623$\pm$0.020     & 0.652$\pm$0.061   & --                    & 1.76$\pm$0.13\\	
6  & --	                      & 0.634$\pm$0.039     & 0.58$\pm$0.10     & --                    & 2.29$\pm$0.30\\	
7  & --	               	      & 2.04$\pm$0.80       & 0.52$\pm$0.17     & --                    & 4.29$\pm$0.59\\
8  & --	                      & 2.17$\pm$0.44       & 0.445$\pm$0.083   & --                    & 4.23$\pm$0.21\\	
9  & 0.90$^{+0.32}_{-0.14}$   & 2.64$\pm$0.38       & 0.256$\pm$0.062   & --                    & 5.30$\pm$0.22\\
10 & 1.60$\pm$0.21            & 2.52$\pm$0.44       & 0.545$\pm$0.074   & 5.4$^{+3.6}_{-2.0}$   & 6.29$\pm$0.29\\	
11 & 1.64$^{+0.25}_{-0.17}$   & 3.8$^{+1.1}_{-0.7}$ & 0.603$\pm$0.058   & 6.5$^{+4.0}_{-1.8}$   & 6.62$\pm$0.22\\	
12 & 1.62$\pm$0.25            & 3.44$\pm$0.88       & 0.762$\pm$0.086   & 4.29$\pm$0.98         & 6.78$\pm$0.38\\	
13 & 0.690$\pm$0.059          & --	            & 1.12$\pm$0.12     & 3.89$\pm$0.68         & 6.04$\pm$0.45\\	
14 & 1.78$\pm$0.18            & --	            & 1.44$\pm$0.22     & 6.21$\pm$0.47         & 6.9$\pm$1.1\\	
15 & 1.99$\pm$0.11            & --	            & 1.53$\pm$0.24     & 7.52$\pm$0.37         & 7.9$\pm$1.0\\	
16 & 2.20$\pm$0.16            & --	            & 1.37$\pm$0.23     & 4.96$\pm$0.20         & 10.5$\pm$2.0\\	
17 & 1.67$\pm$0.28            & --	            & 1.11$\pm$0.46     & 13.2$\pm$1.3          & 10.7$^{+5.5}_{-3.7}$\\
18 & 0.62$\pm$0.18$^{1}$      & --	            & --	        &  --	                &  --\\
19 & 0.68$\pm$0.22$^{1}$      & --	            & --	        &  --	                &  --\\
\hline
\multicolumn{6}{c}{4U 0614+09}\\
\hline
1  & --	                      & 0.504$\pm$0.017     & --	        & 0.057$\pm$0.039$^{1}$ & 0.512$\pm$0.070$^{1}$\\
2  & --	                      & 0.464$\pm$0.0177    & 0.252$\pm$0.058   & --                    & 1.52$\pm$0.21\\
3  & --	                      & 0.555$\pm$0.065     & 0.61$\pm$0.14     & --                    & 1.94$\pm$0.38\\
4  & --	                      & 0.98$\pm$0.22       & 0.36$\pm$0.10     & --                    & 3.66$\pm$0.30\\
5  & 0.81$\pm$0.17            & 6.2$^{+4.3}_{-2.4}$ & 0.217$\pm$0.067   & --                    & 4.34$\pm$0.29\\
6  & 1.04$\pm$0.14            & 6$^{+14}_{-3}$      & 0.576$\pm$0.079   & 5.9$^{+3.1}_{-1.6}$   & 5.17$\pm$0.34\\
7  & 1.44$\pm$0.19            & --	            & 0.79$\pm$0.11     & 13$^{+10}_{-3}$       & 11.1$\pm$1.2\\
8  & 1.59$\pm$0.15            & --	            & 1.27$\pm$0.18     & 10.5$\pm$1.2          & 7.02$\pm$0.86\\
9  & 0.635$\pm$0.059$^{1}$    & --	            & 2.79$\pm$0.27     & --                    & 10.9$\pm$2.9\\
\enddata

\tablecomments{Listed are the $Q$ values ($\equiv \nu_{\rm 0}/2\Delta$) of 
the different Lorentzians described in \S 3. The quoted 
errors in $Q$ use $\Delta\chi^{2}$ = 1.0.\\
$^{1}$ The identification of this Lorentzian is not clear, see \S 3. 
}

\end{deluxetable}
\end{center}

\begin{center}
\begin{deluxetable}{lccccccc}
\tabletypesize{\footnotesize}
\tablewidth{0pt}
\tablecaption{Integrated fractional rms of the multi--Lorentzian fit for 
4U 1728--34 \& 4U 0614+09\label{tbl.rms}} 
\tablehead{ 
\colhead{Interval} & \colhead{zc Lorentzian} & 
\colhead{VLF Lorentzian} & \colhead{LF Lorentzian} & 
\colhead{hHz Lorentzian} & 
\colhead{Lower kHz QPO} & \colhead{Upper kHz QPO} \\
\colhead{Number} & \colhead{rms (\%)} & 
\colhead{rms (\%)} & \colhead{rms (\%)} & 
\colhead{rms (\%)} & \colhead{rms (\%)} & 
\colhead{rms (\%)}
}

\startdata
\multicolumn{7}{c}{4U 1728--34}\\
\hline
1  & 13.32$\pm$0.17   & --	            & 14.43$\pm$0.30      & 13.7$\pm$1.2            & --            & 9.1$\pm$1.4\\
2  & 13.11$\pm$0.14   & --	            & 14.46$\pm$0.25      & 13.17$\pm$0.80          & --            & 9.64$\pm$0.89\\
3  & 13.08$\pm$0.19   & --	            & 14.56$\pm$0.26      & 11.8$\pm$1.3            & --            & 10.72$\pm$1.28\\
4  & 13.82$\pm$0.10   & --	            & 13.47$\pm$0.15      & 10.67$\pm$0.34          & --            & 12.10$\pm$0.32\\	
5  & 13.878$\pm$0.093 & --	            & 13.17$\pm$0.14      & 11.13$\pm$0.35          & --            & 11.82$\pm$0.34\\	
6  & 13.89$\pm$0.17   & --	            & 13.00$\pm$0.27      & 11.71$\pm$0.64          & --            & 10.90$\pm$0.61\\	
7  & 15.47$\pm$0.49   & --	            & 5.8$^{+1.4}_{-0.9}$ & 12.47$\pm$0.74          & --            & 12.07$\pm$0.56\\
8  & 14.93$\pm$0.24   & --	            & 4.80$\pm$0.51       & 10.74$\pm$0.36          & --            & 12.38$\pm$0.21\\	
9  & 11.2$\pm$2.6     & 6.2$^{+4.0}_{-1.7}$ & 4.83$\pm$0.39       & 12.23$^{+0.39}_{-0.78}$ & --            & 11.95$\pm$0.17\\
10 & 11.68$\pm$0.82   & 5.03$\pm$0.75       & 3.99$\pm$0.42       & 10.05$\pm$0.36          & 3.02$\pm$0.52 & 10.81$\pm$0.19\\	
11 & 11.22$\pm$0.69   & 5.18$\pm$0.59       & 2.87$\pm$0.38       & 9.19$\pm$0.36           & 2.91$\pm$0.34 & 10.43$\pm$0.13\\	
12 & 9.90$\pm$0.77    & 5.89$\pm$0.67       & 3.09$\pm$0.40       & 8.93$\pm$0.37           & 4.50$\pm$0.37 & 9.96$\pm$0.22\\	
13 & 6.12$\pm$0.38    & 8.46$\pm$0.33       & --	          & 6.83$\pm$0.23           & 5.45$\pm$0.38 & 8.20$\pm$0.24\\	
14 & 5.73$\pm$0.26    & 5.41$\pm$0.25       & --	          & 4.68$\pm$0.24           & 6.92$\pm$0.21 & 5.33$\pm$0.32\\	
15 & 4.36$\pm$0.13    & 5.16$\pm$0.12       & --	          & 3.91$\pm$0.19           & 7.01$\pm$0.12 & 4.50$\pm$0.21\\	
16 & 3.92$\pm$0.16    & 4.73$\pm$0.13       & --	          & 3.60$\pm$0.20           & 7.61$\pm$0.13 & 3.15$\pm$0.26\\	
17 & 2.91$\pm$0.41    & 4.49$\pm$0.29       & --	          & 3.82$\pm$0.46           & 5.71$\pm$0.22 & 3.41$\pm$0.50\\
18 & --		      & 2.77$\pm$0.18$^{1}$ & --	          & --	                    &  --	    &  --\\
19 & --		      & 2.21$\pm$0.20$^{1}$ & --	          & --	                    &  --	    &  --\\
\hline
\multicolumn{7}{c}{4U 0614+09}\\
\hline
1  & 17.95$\pm$0.11      & --	               & 20.18$\pm$0.23         &  --	         & 21.43$\pm$0.37$^{1}$   & 16.55$\pm$0.52$^{1}$\\
2  & 16.54$\pm$0.11      & --	               & 19.20$\pm$0.18         & 17.69$\pm$0.83 & --                     & 12.9$\pm$1.0\\
3  & 17.45$\pm$0.38      & --	               & 17.28$\pm$0.64         & 17.6$\pm$1.2   & --                     & 16.1$\pm$1.2\\
4  & 17.45$\pm$0.63      & --	               & 10.8$\pm$1.3           & 18.84$\pm$0.80 & --                     & 16.90$\pm$0.53\\
5  & 7.8$^{+1.8}_{-1.1}$ & 11.2$\pm$1.2        & 3.57$^{+0.89}_{-0.53}$ & 19.97$\pm$0.53 & --                     & 16.34$\pm$0.37\\
6  & 8.5$^{+1.0}_{-0.7}$ & 10.64$\pm$0.72      & 3.3$^{+1.2}_{-0.5}$    & 16.34$\pm$0.48 & 5.55$\pm$0.67          & 14.42$\pm$0.34\\
7  & 7.8$^{+1.0}_{-0.6}$ & 9.54$\pm$0.62       & --	                   & 12.98$\pm$0.46 & 9.45$^{+0.92}_{-0.31}$ & 10.66$\pm$0.42\\
8  & 5.50$\pm$0.77       & 8.32$\pm$0.39       & --	                   & 9.07$\pm$0.40  & 9.21$\pm$0.35          & 10.31$\pm$0.43\\
9  & --                  & 6.09$\pm$0.16$^{1}$ & --	                   & 6.33$\pm$0.21  & --                     & 5.27$\pm$0.46\\
\enddata

\tablecomments{Listed are the values of the integrated fractional rms
(over the full PCA energy band) of the different Lorentzians described 
in \S 3. The quoted errors in the rms use $\Delta\chi^{2}$ = 1.0.\\
$^{1}$ The identification of this Lorentzian is not clear, see \S 3. 
}

\end{deluxetable}
\end{center}

\onecolumn

\begin{figure}
\figurenum{1}
\epsscale{0.45}
\begin{tabular}{ccc}
\plotone{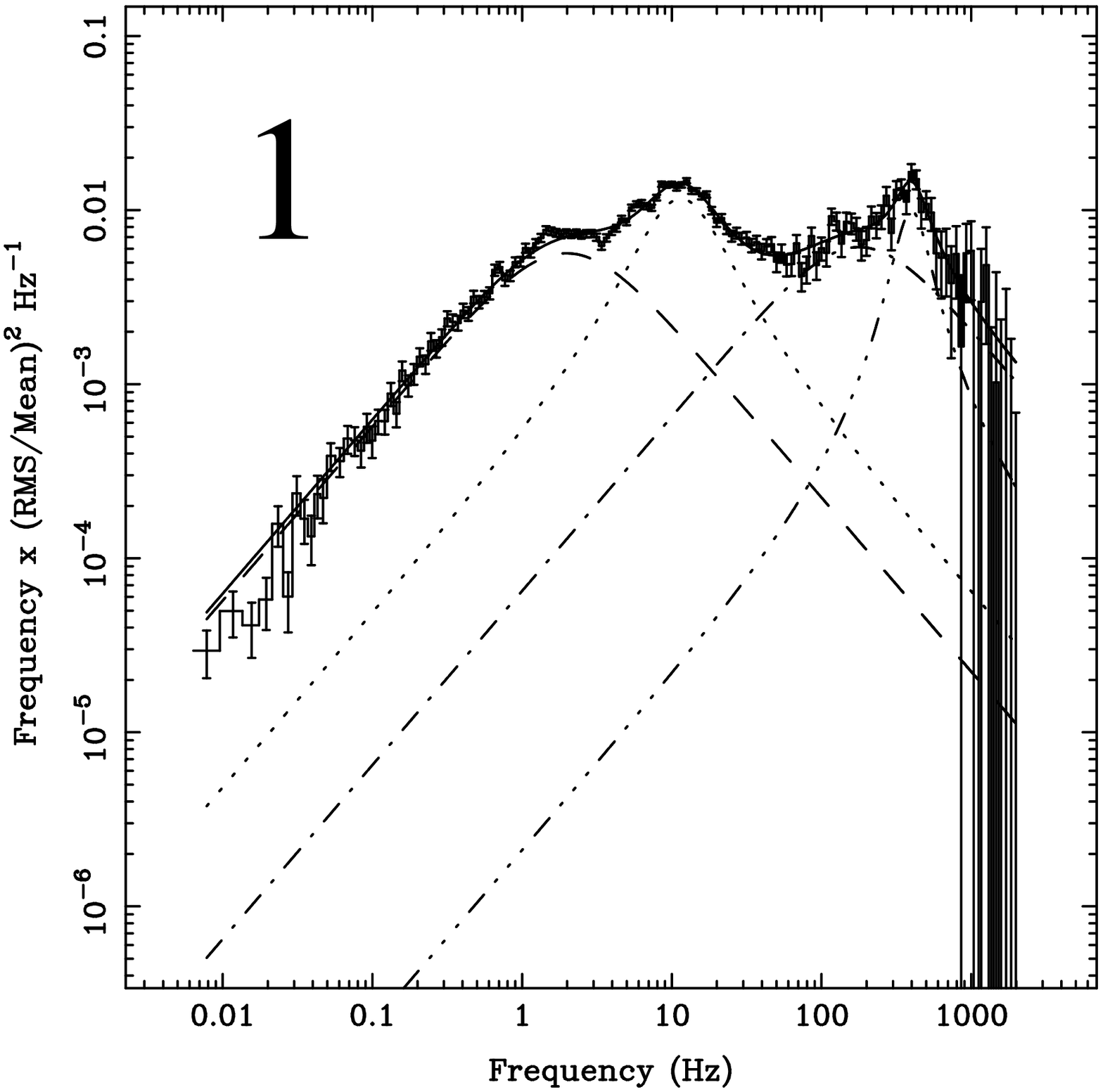} & 
\plotone{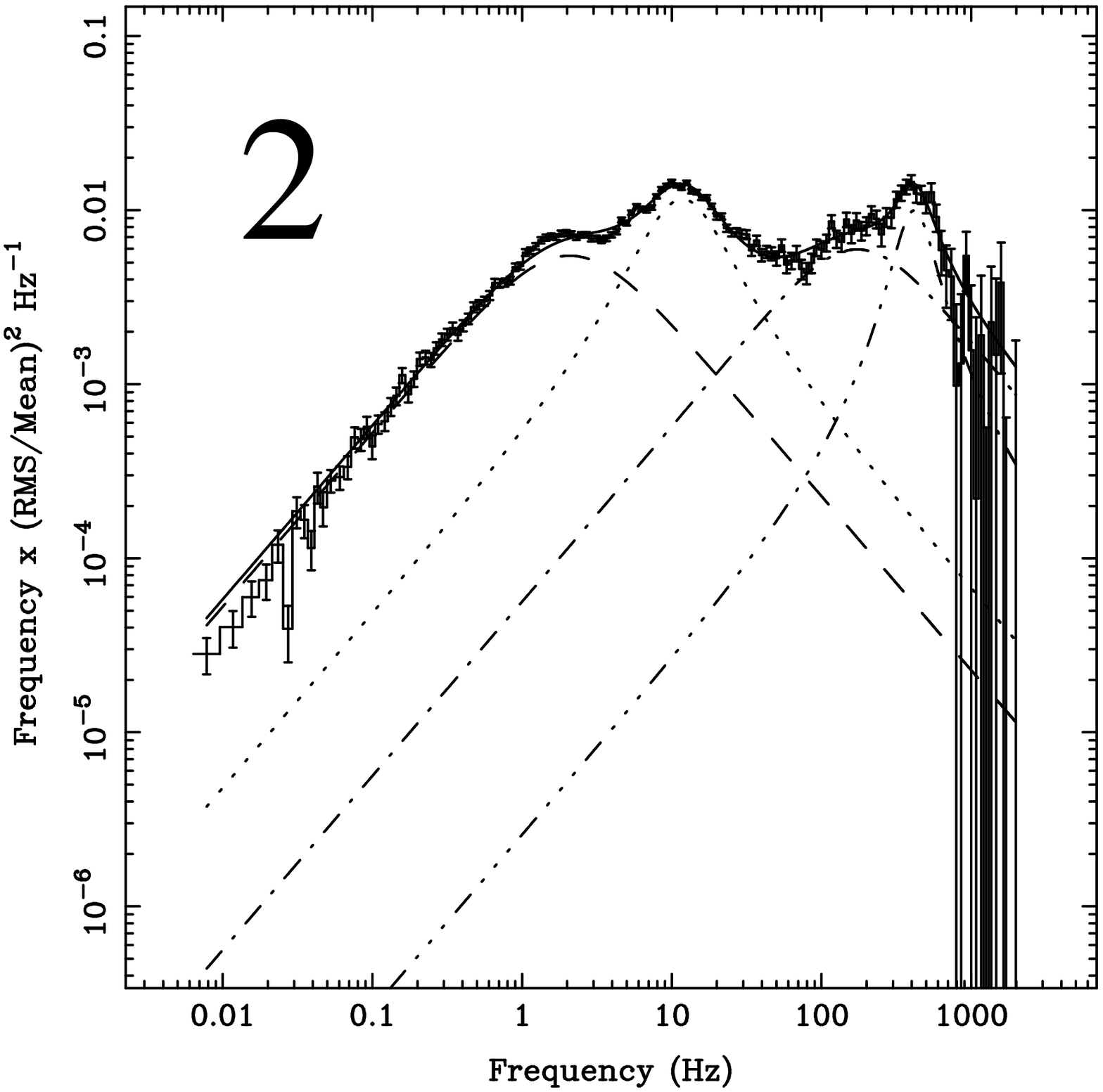} & \\
\plotone{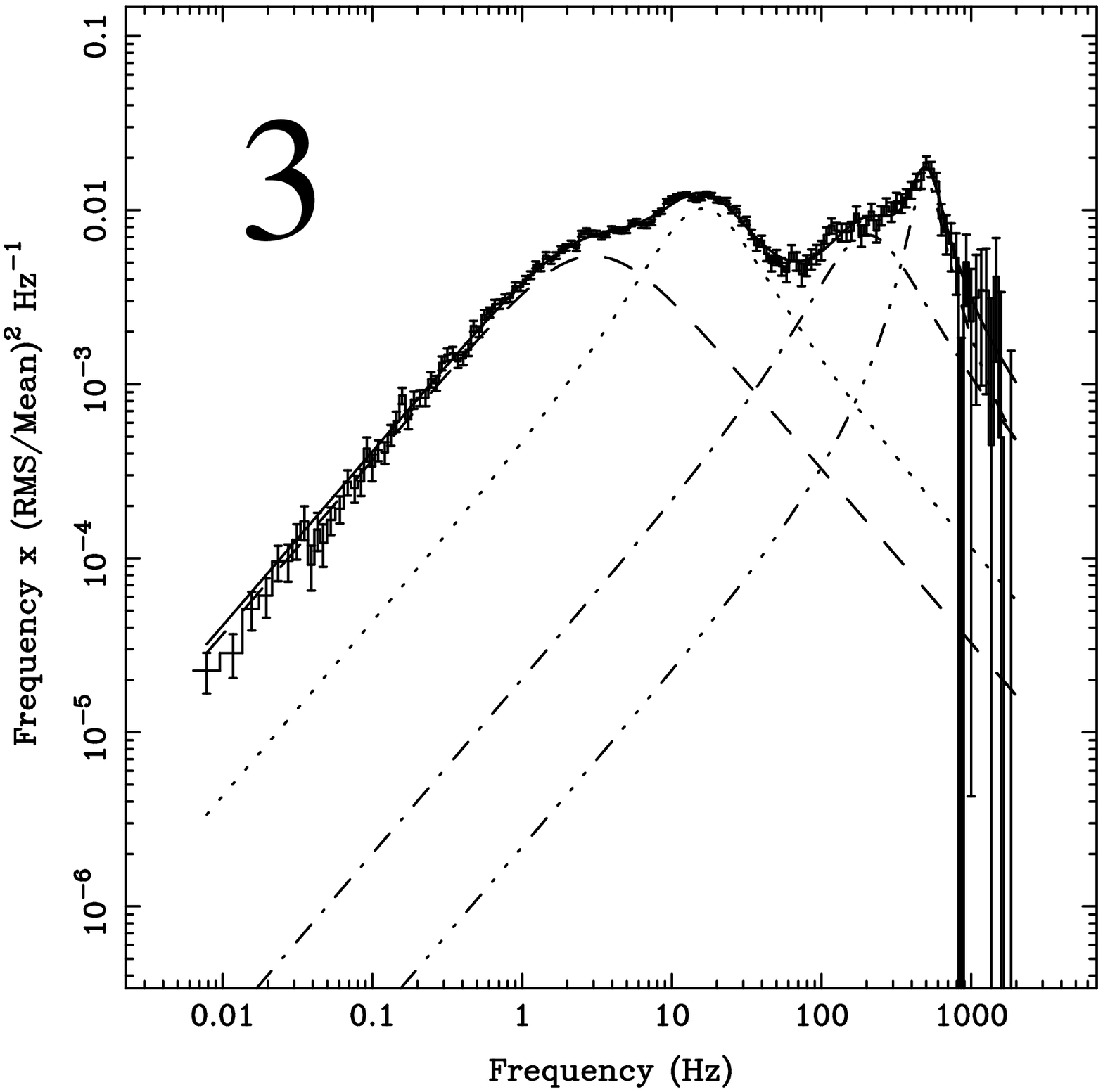} & 
\plotone{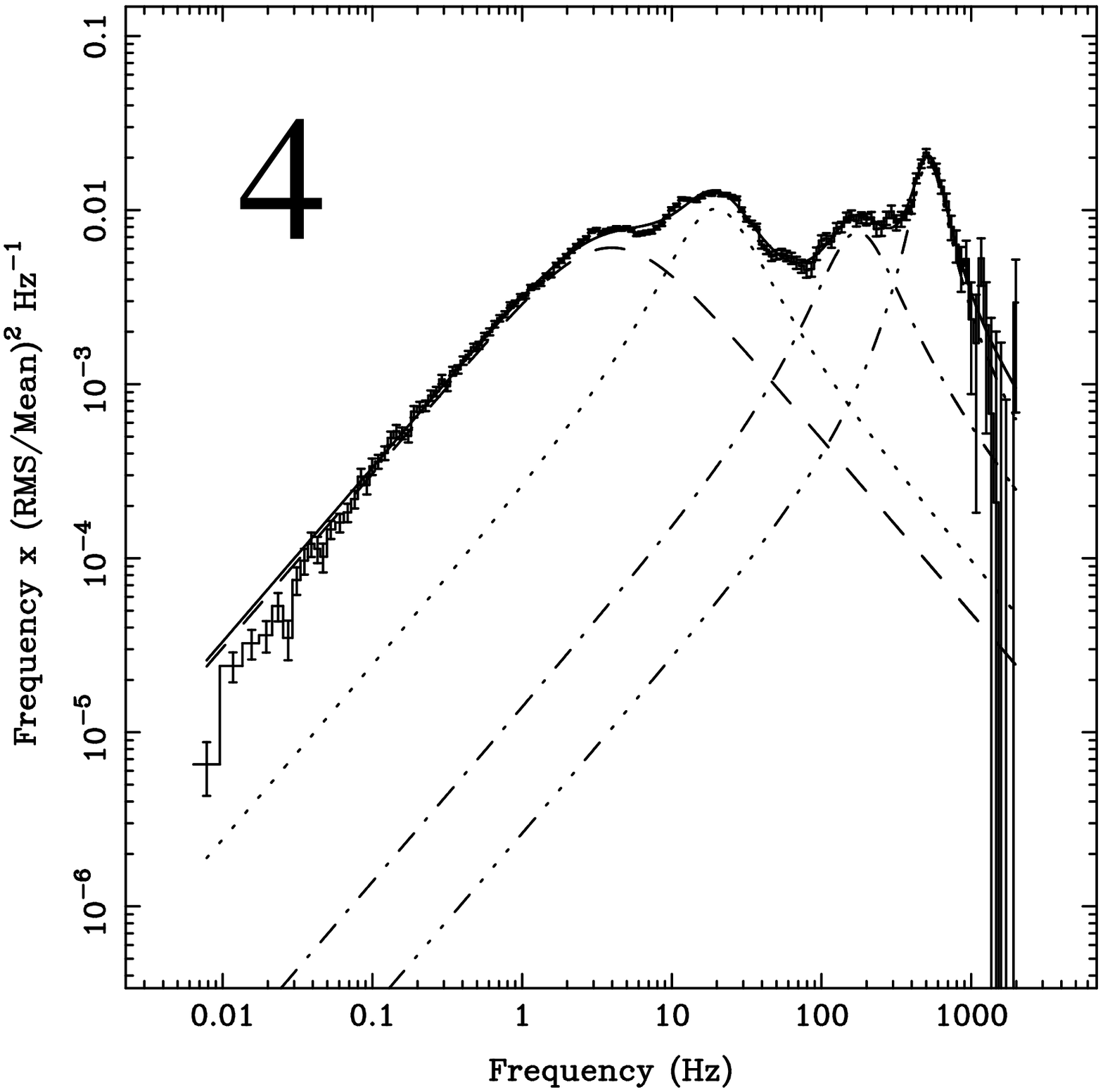} & \\
\end{tabular}
\caption{Power spectra and fit functions in the power spectral density 
times frequency representation (see \S 2) for 4U 1728--34. The different 
lines mark the individual Lorentzian components of the fit. The dashed 
lines mark both the BLN and the very low--frequency Lorentzian (\S 3.1), the dotted lines the 
low--frequency Lorentzian (\S 3.2), the dash--dotted line the hectohertz 
Lorentzian (\S 3.3) and the dash--dot--dot--dotted line both kilohertz QPOs 
(\S 3.4). In intervals 14--19 also a power--law is included to fit the VLFN. 
Interval numbers are indicated.}
\label{fig.powspec_1728}
\end{figure}
\clearpage

\begin{figure}
\figurenum{1}
\epsscale{0.45}
\begin{tabular}{ccc}
\plotone{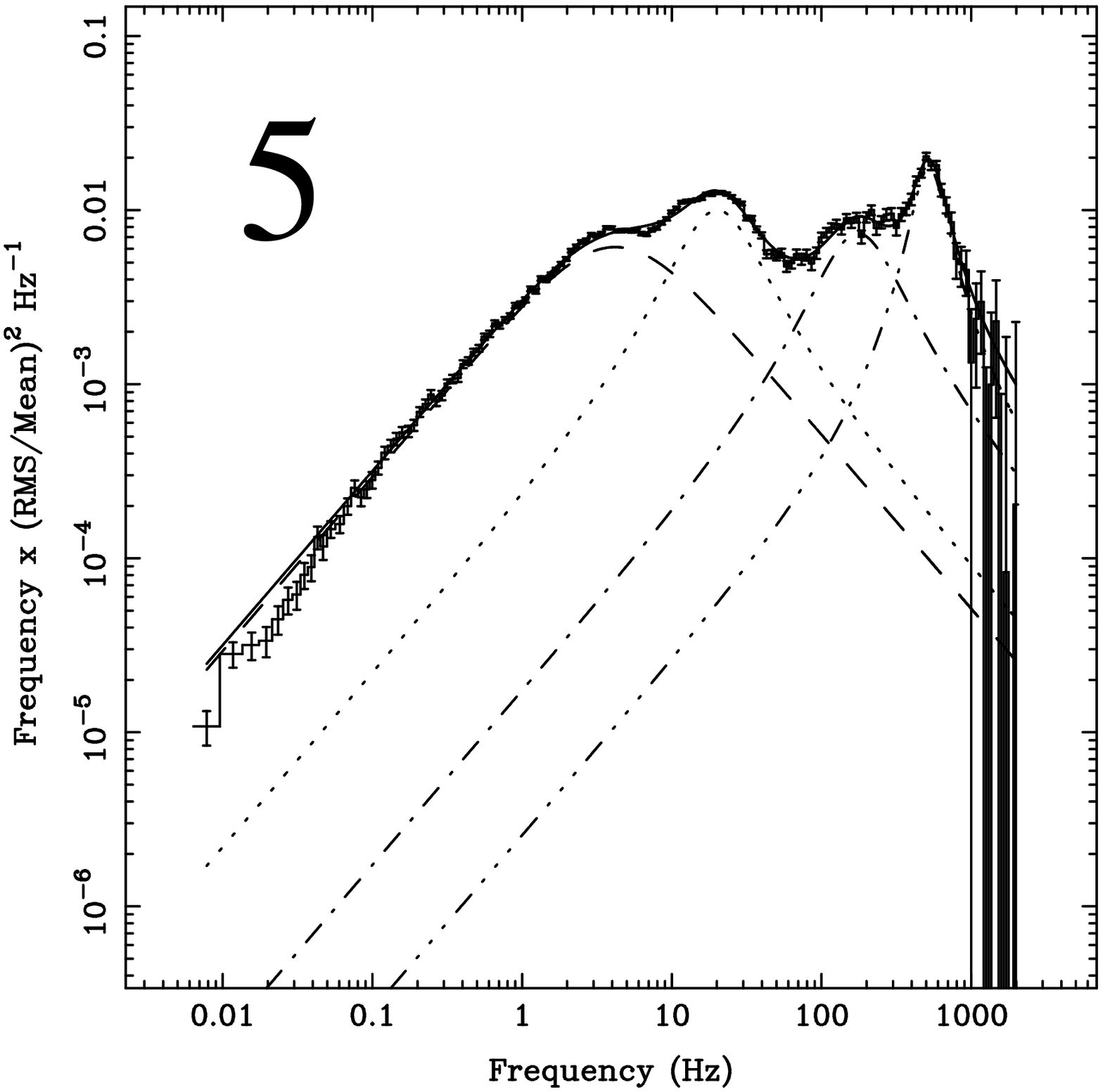} & 
\plotone{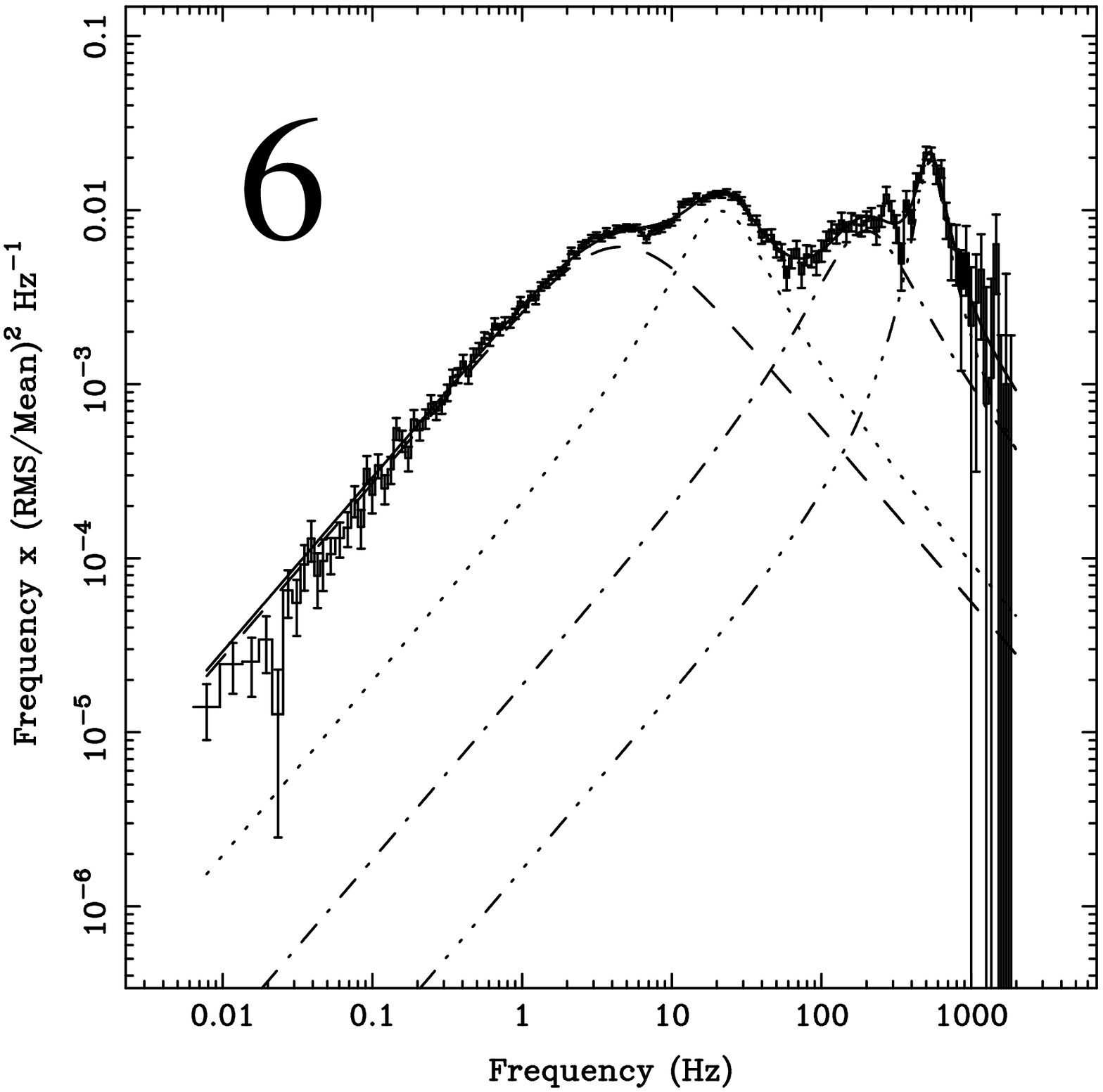} & \\
\plotone{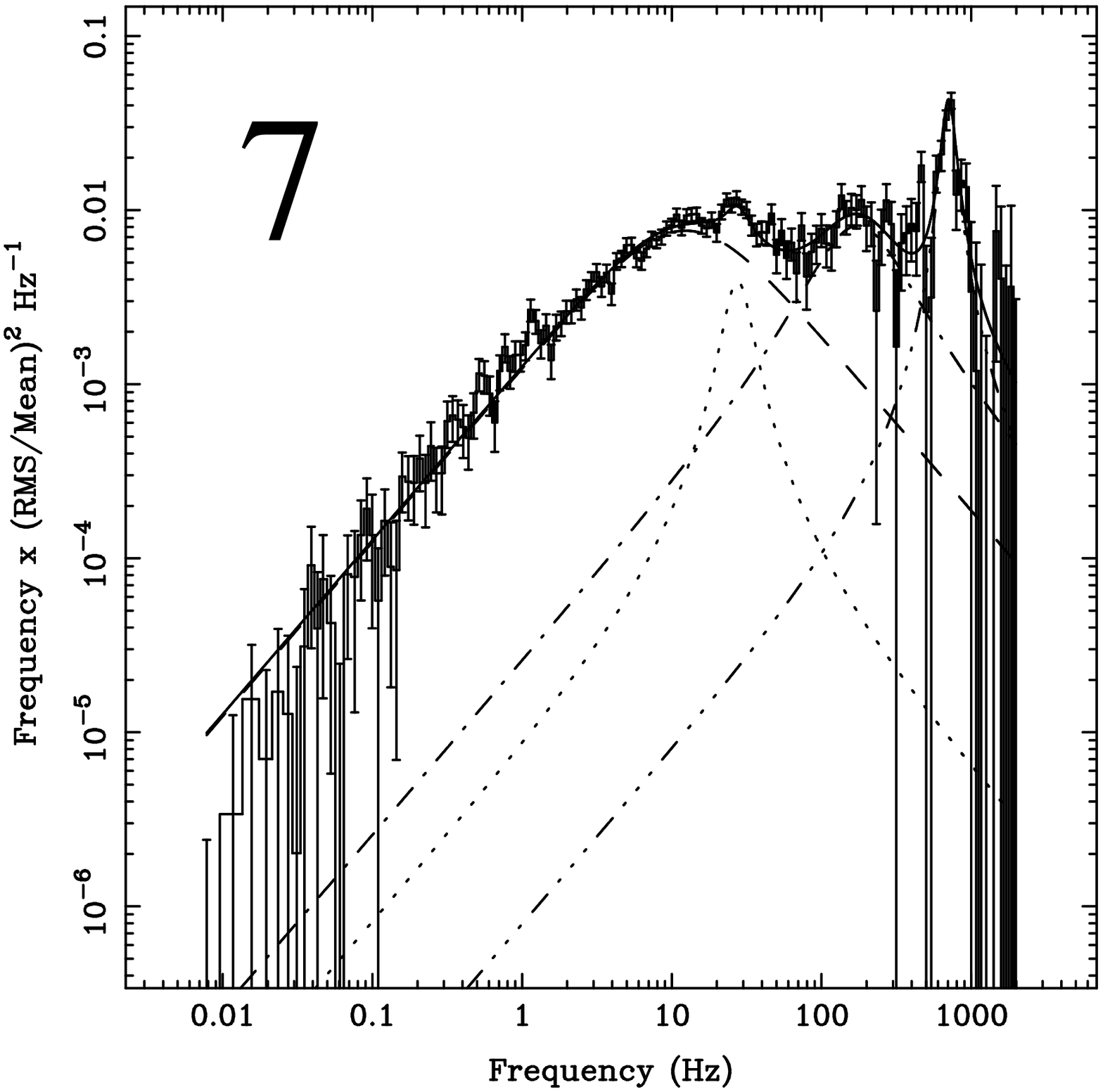} & 
\plotone{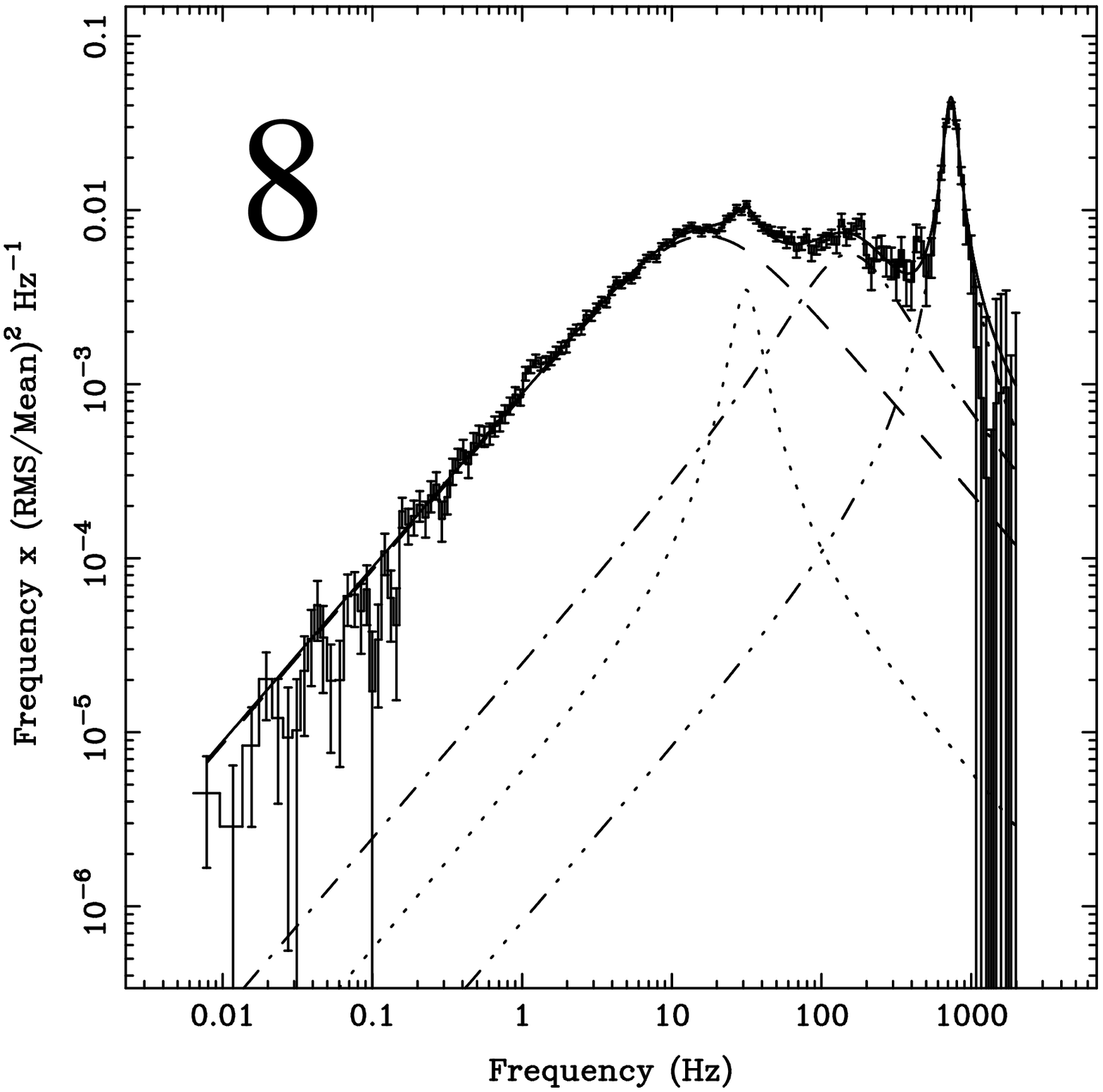} & \\
\end{tabular}
\caption{Continued}
\end{figure}
\clearpage

\begin{figure}
\figurenum{1}
\epsscale{0.45}
\begin{tabular}{ccc}
\plotone{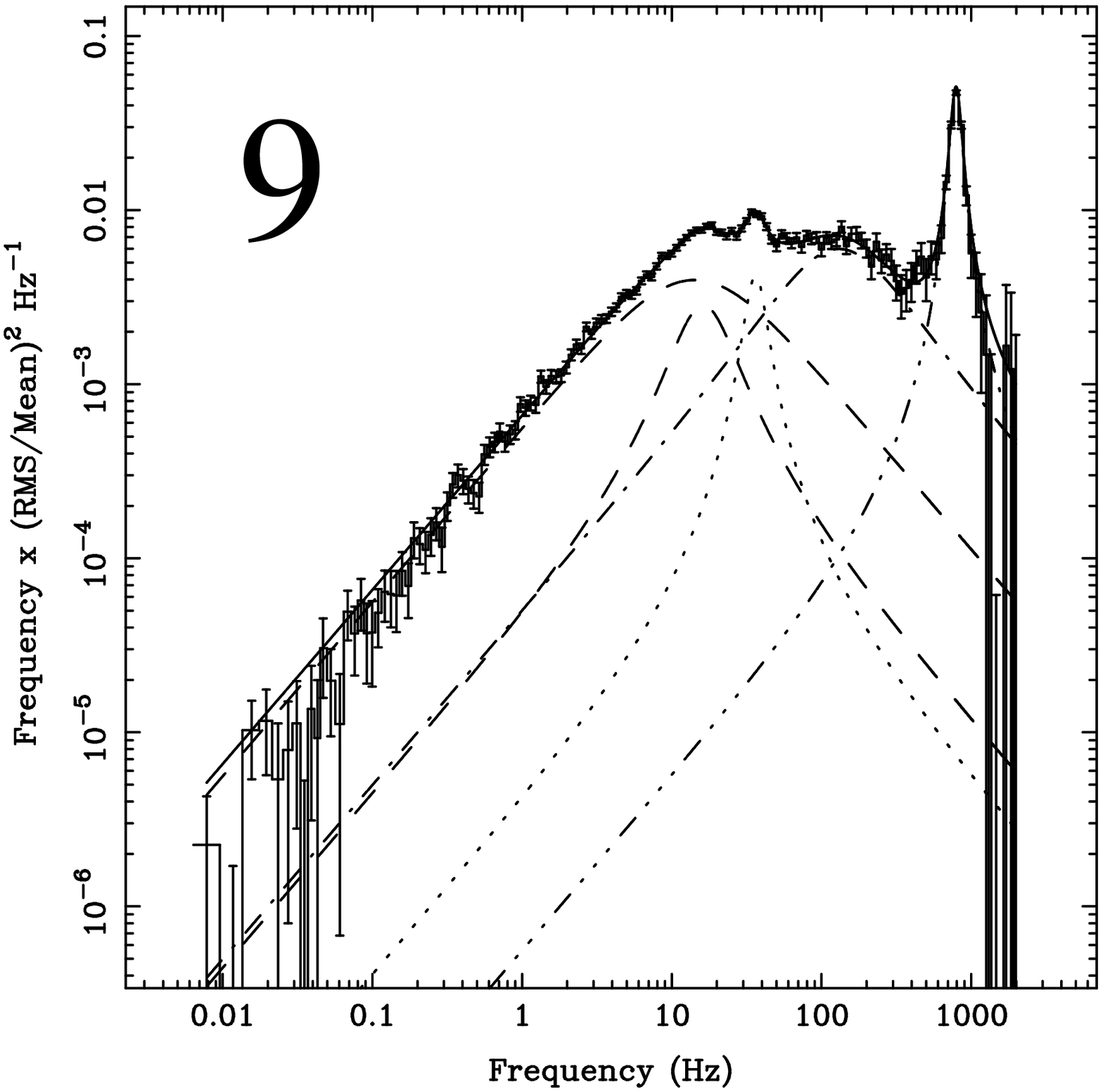} & 
\plotone{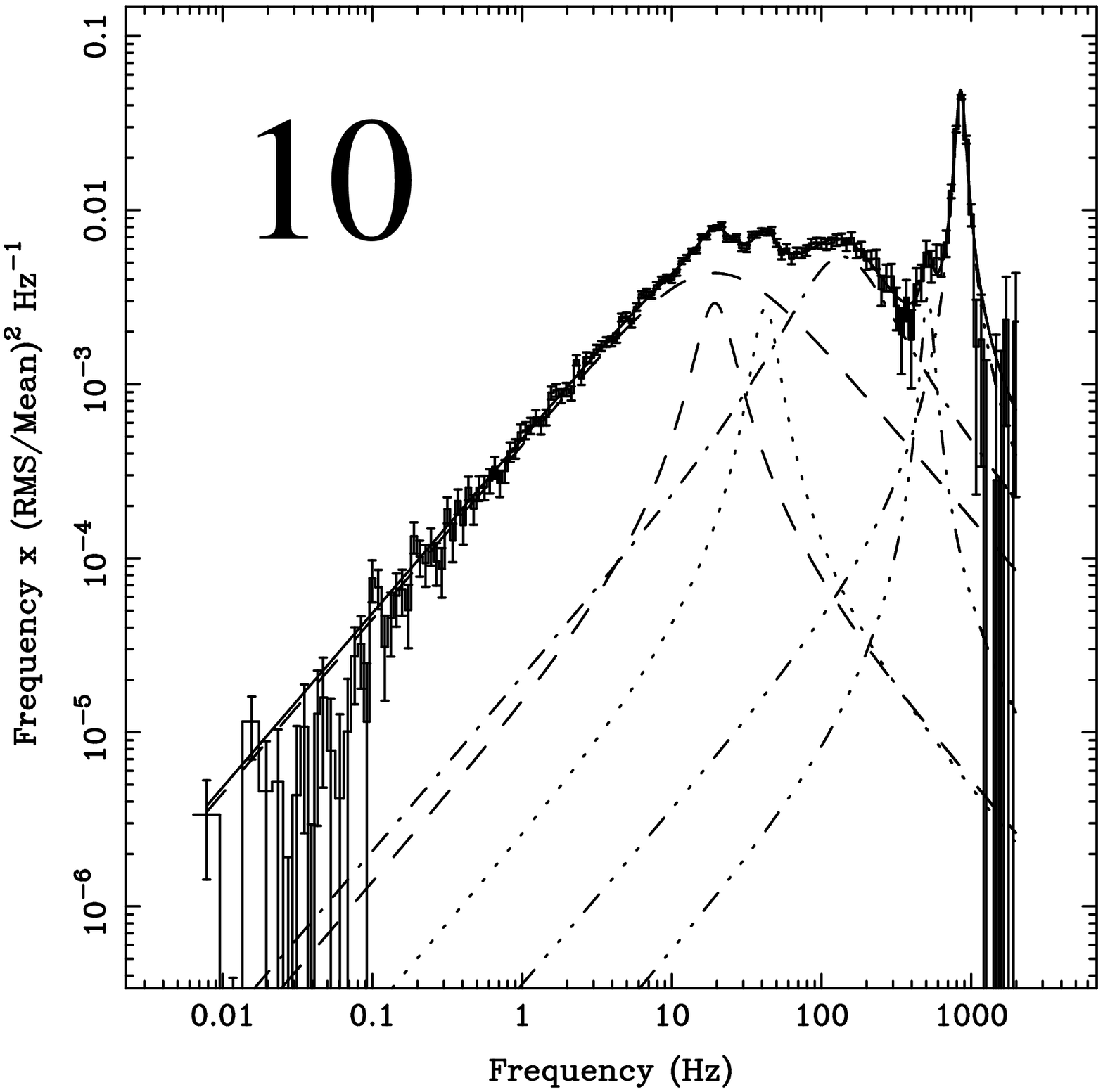} & \\
\plotone{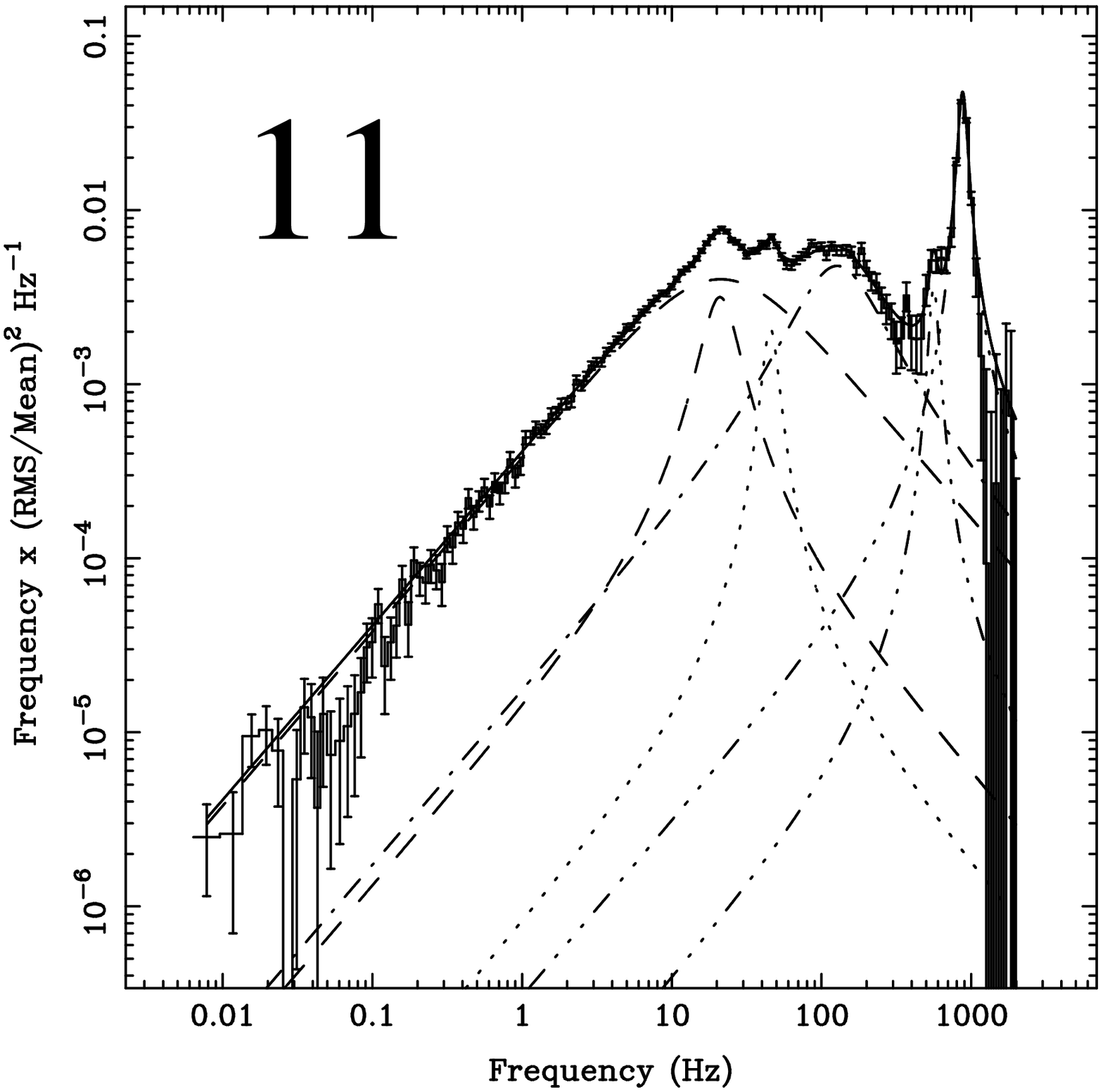} & 
\plotone{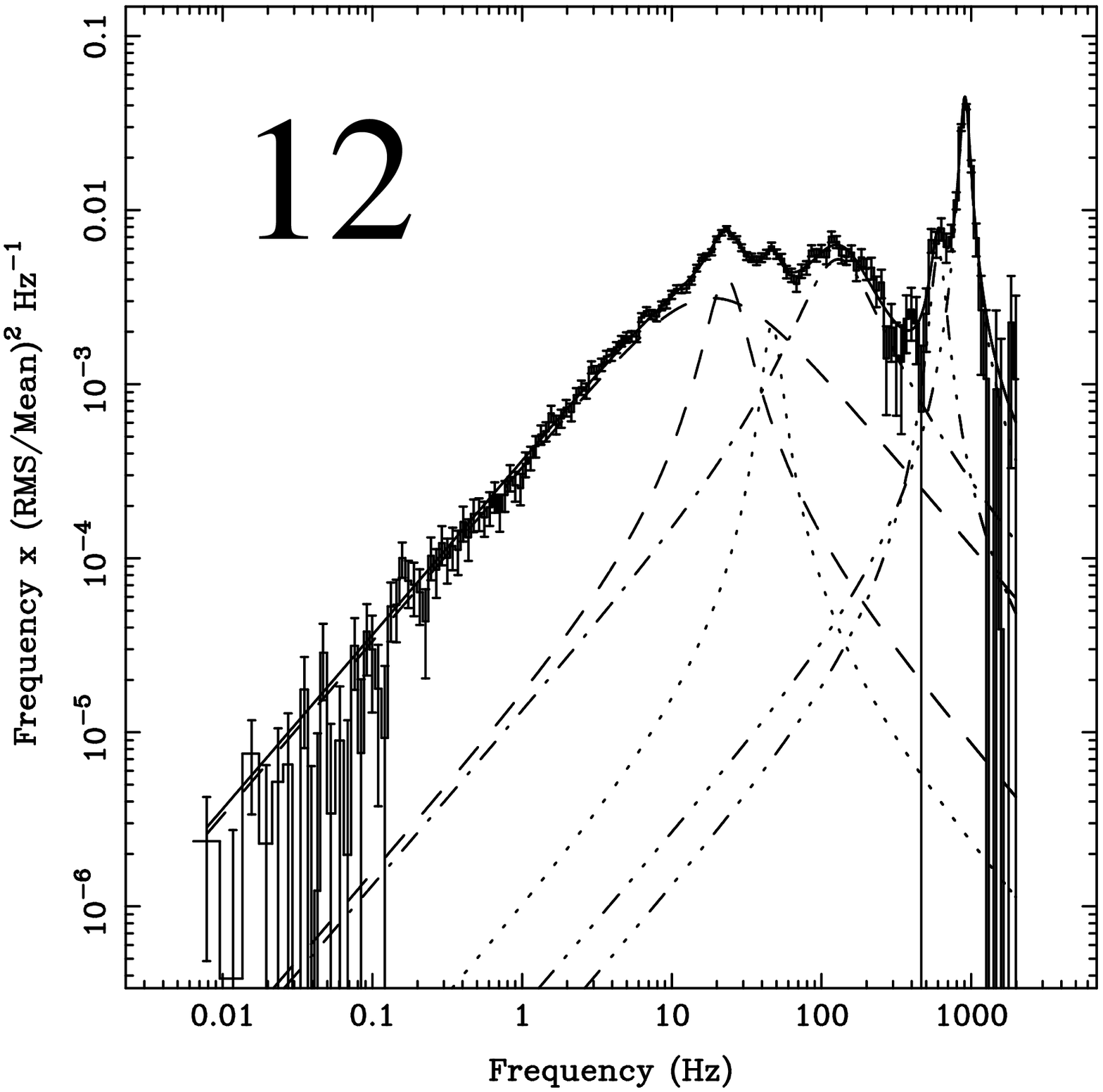} & \\
\end{tabular}
\caption{Continued}
\end{figure}
\clearpage

\begin{figure}
\figurenum{1}
\epsscale{0.45}
\begin{tabular}{ccc}
\plotone{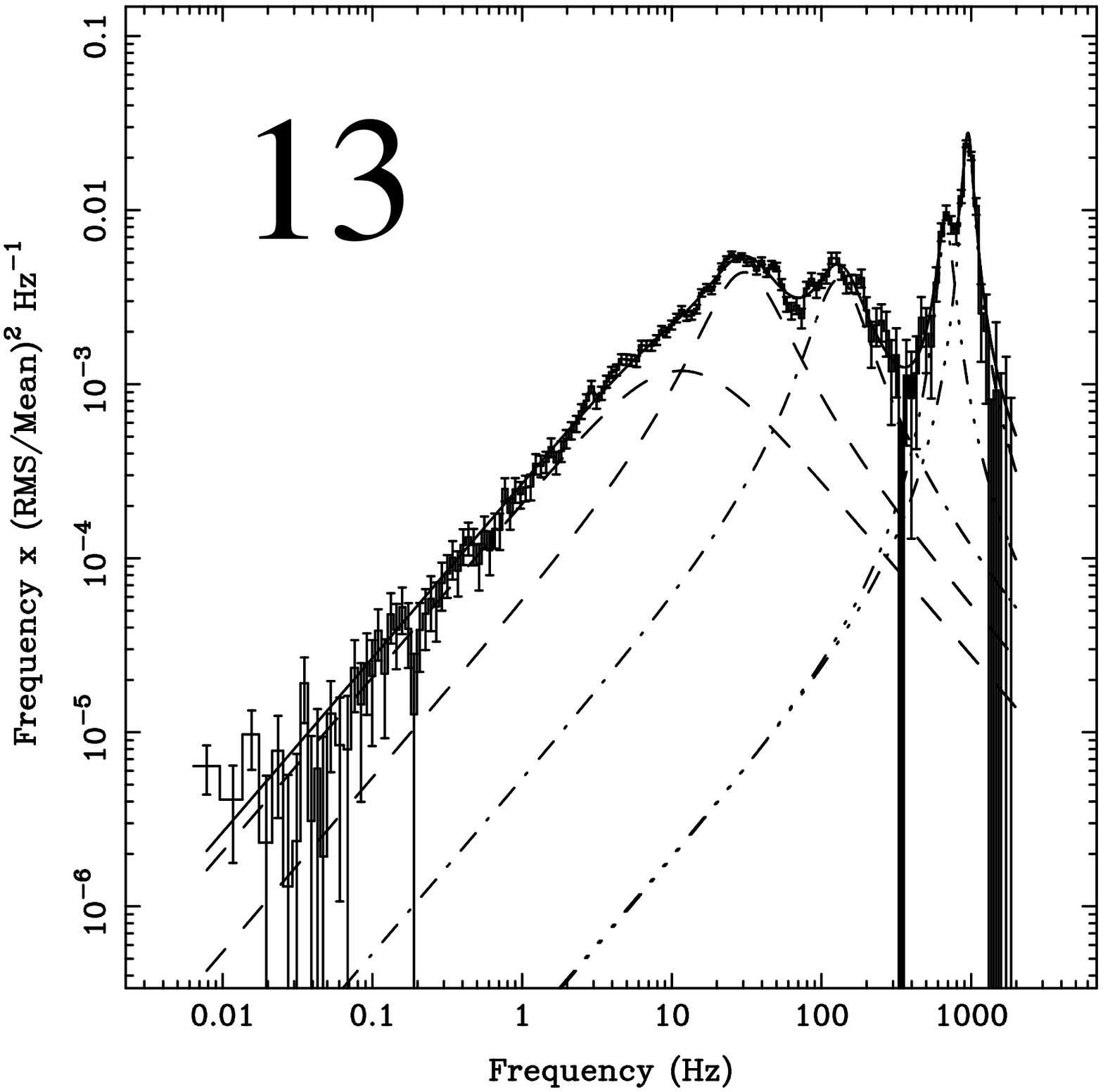} & 
\plotone{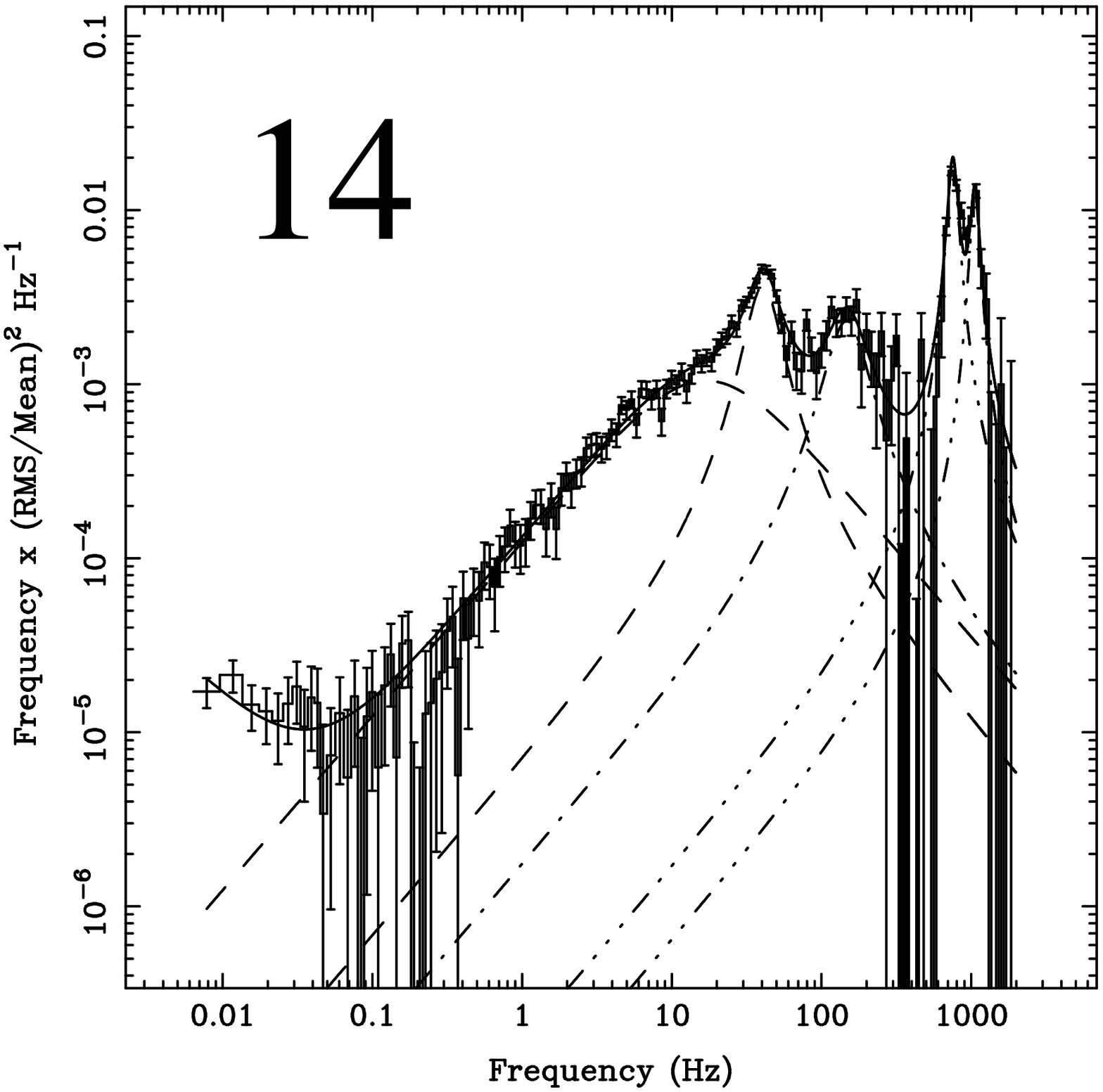} & \\
\plotone{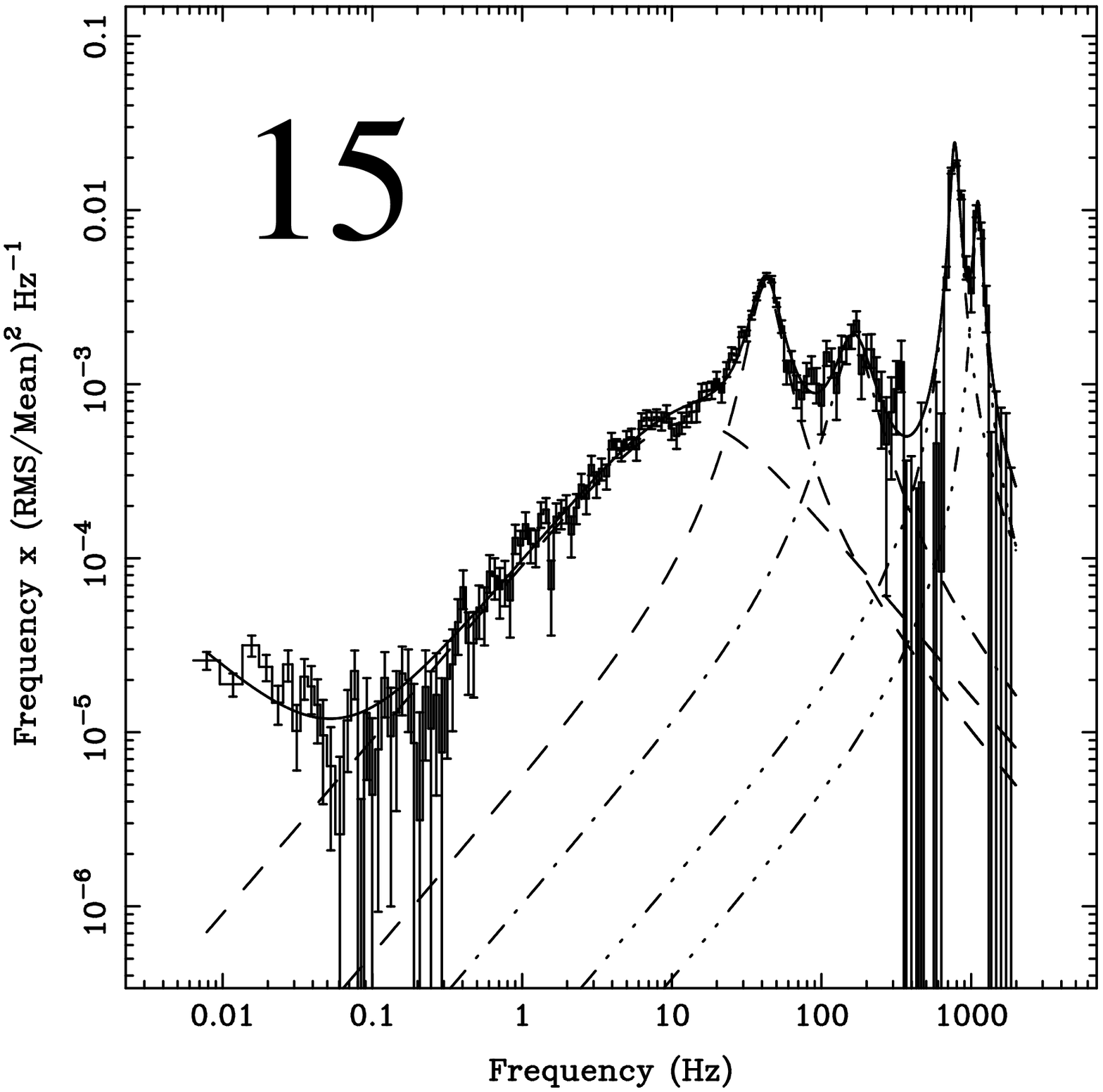} & 
\plotone{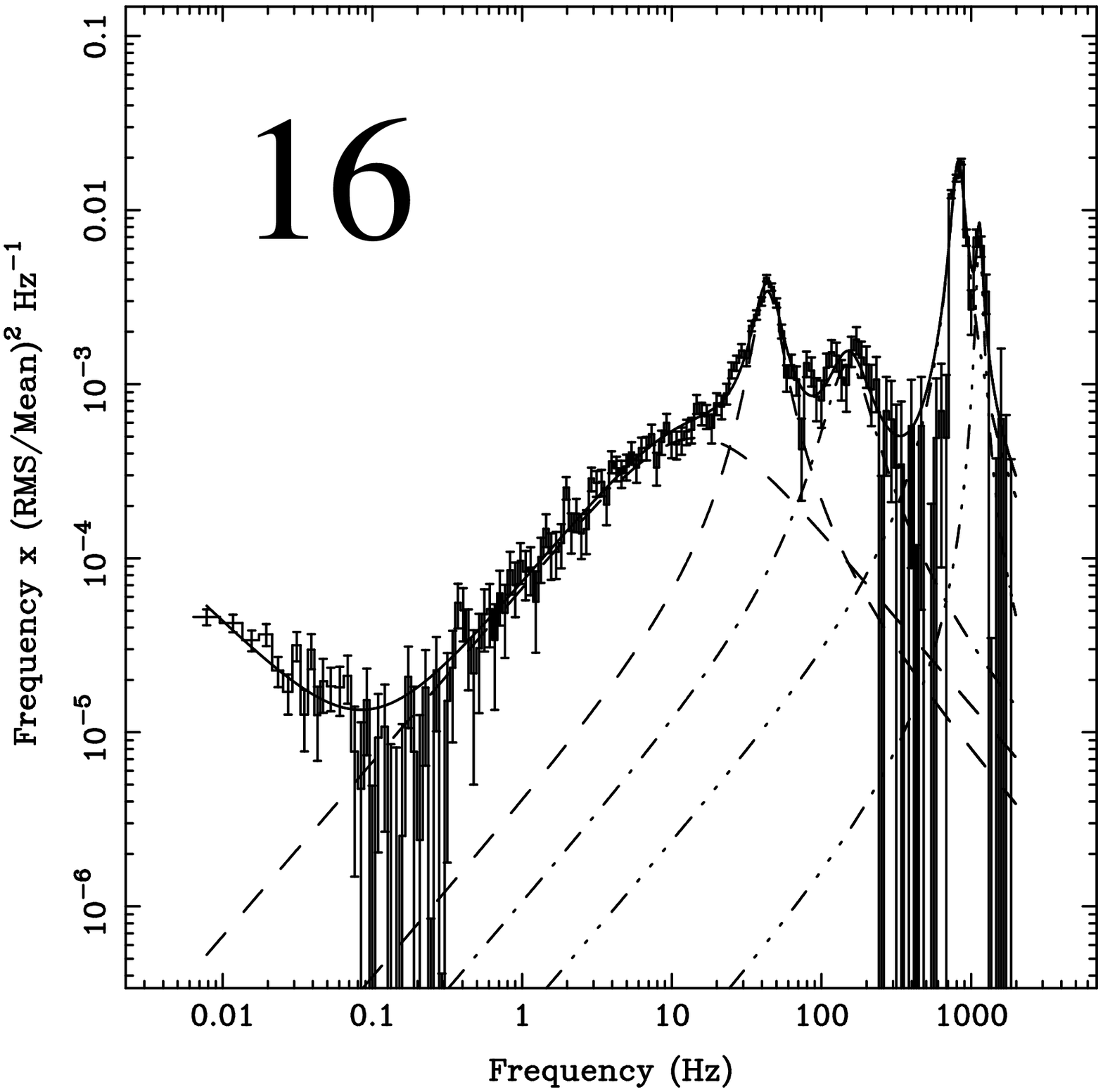} & \\
\end{tabular}
\caption{Continued}
\end{figure}
\clearpage

\begin{figure}
\figurenum{1}
\epsscale{0.45}
\begin{tabular}{ccc}
\plotone{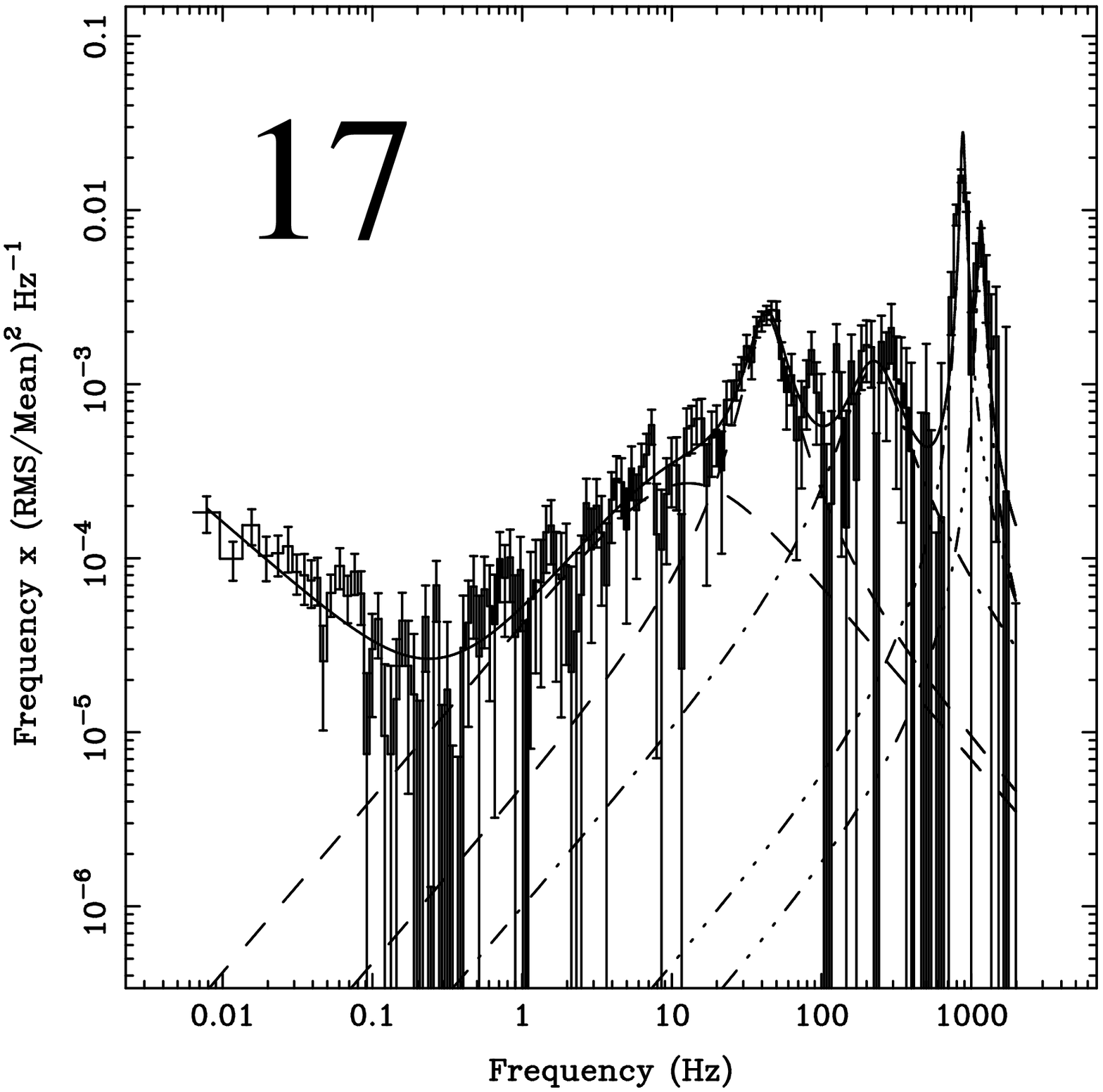} & 
\plotone{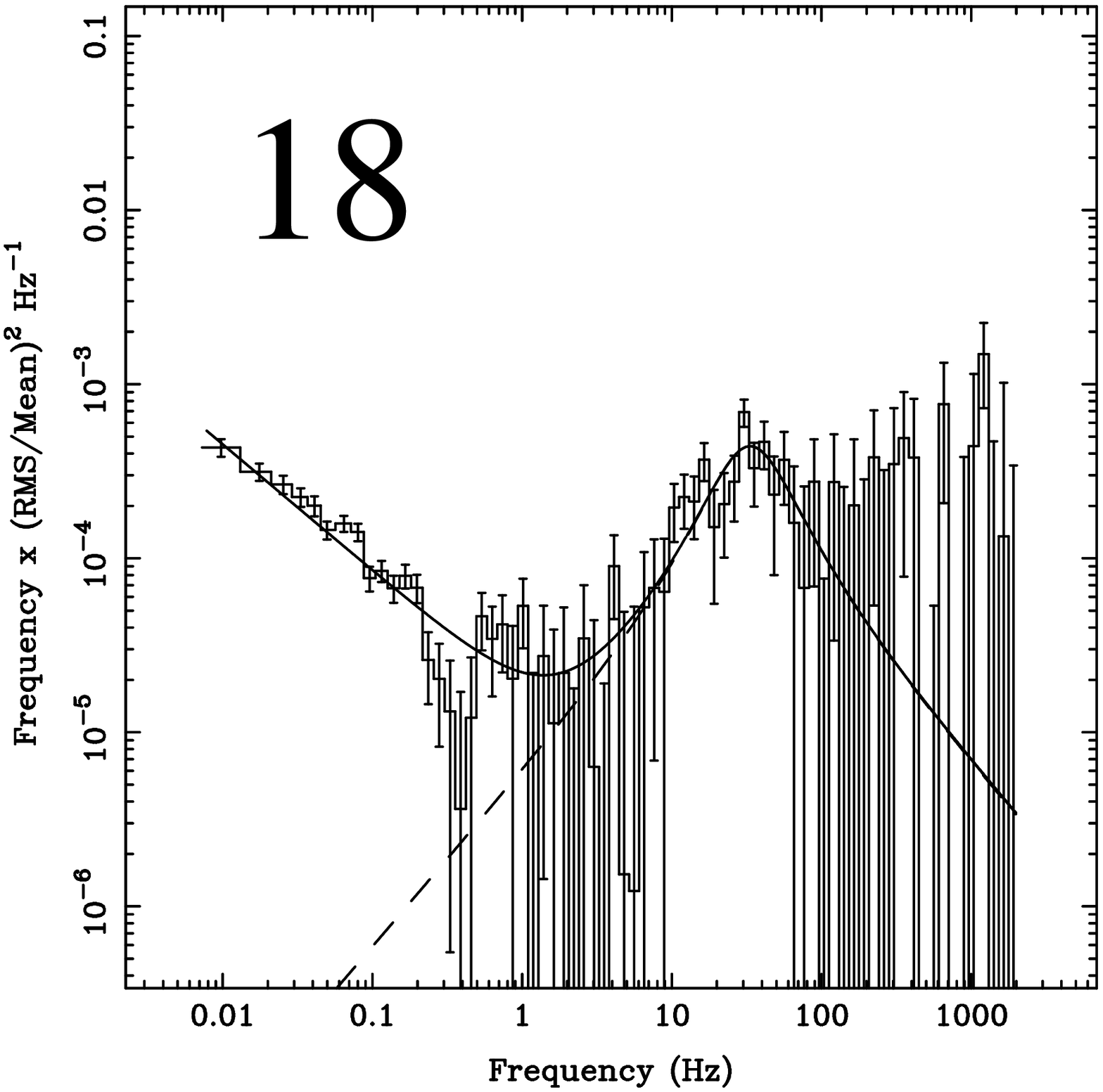} & \\
\plotone{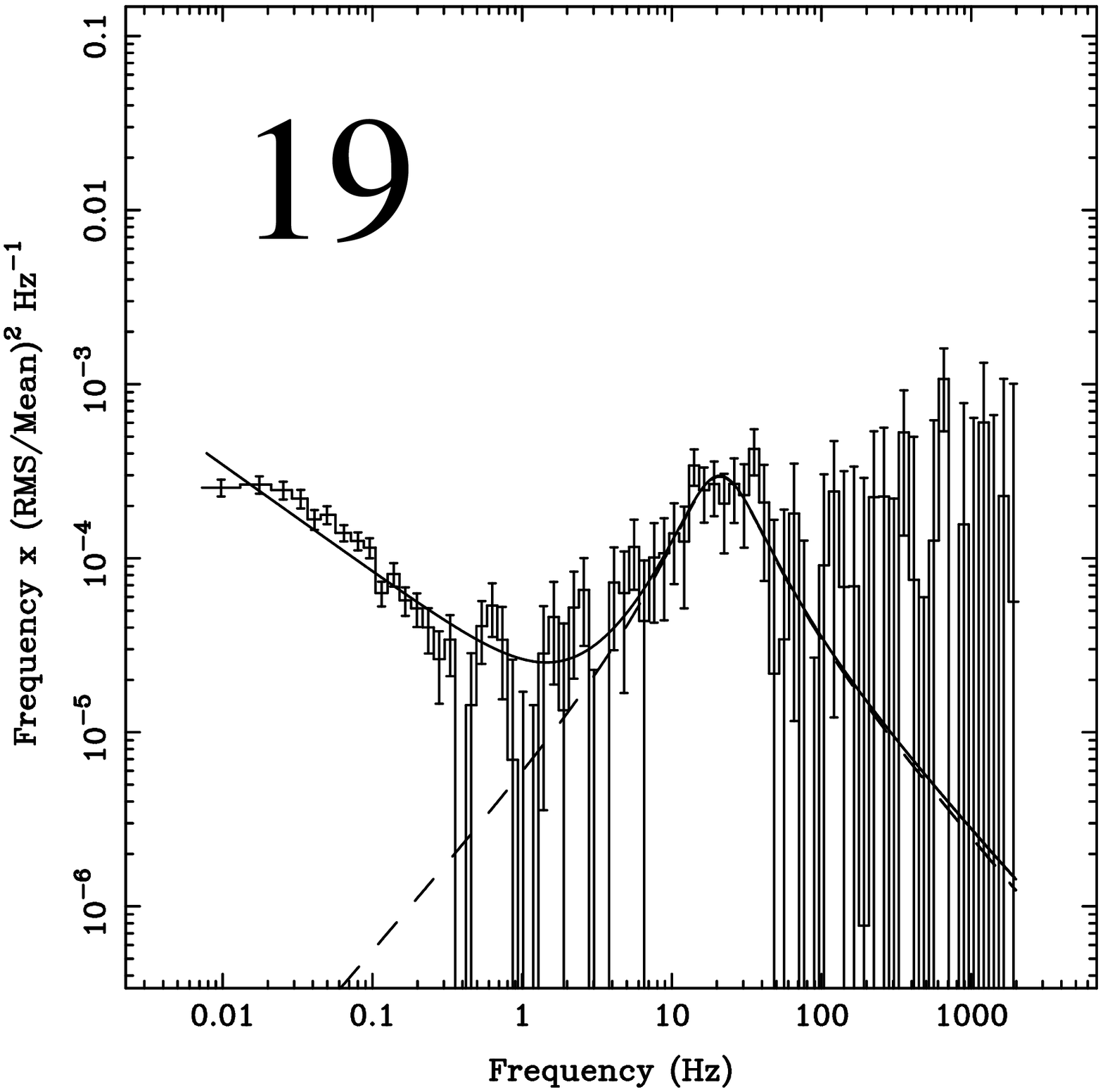} & & \\
\end{tabular}
\caption{Continued}
\end{figure}
\clearpage

\begin{figure}
\figurenum{2}
\epsscale{0.45}
\begin{tabular}{ccc} 
\plotone{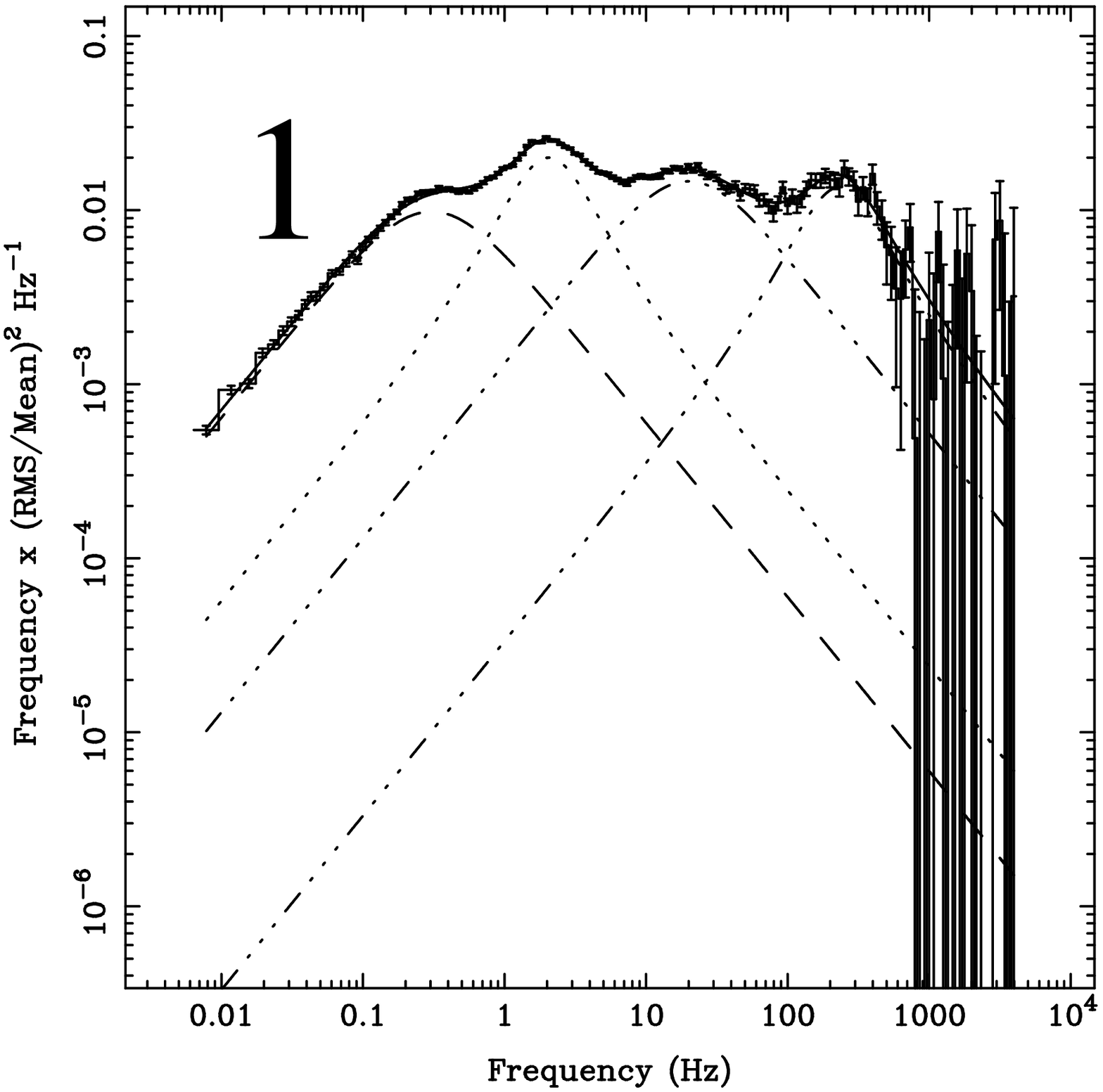} & 
\plotone{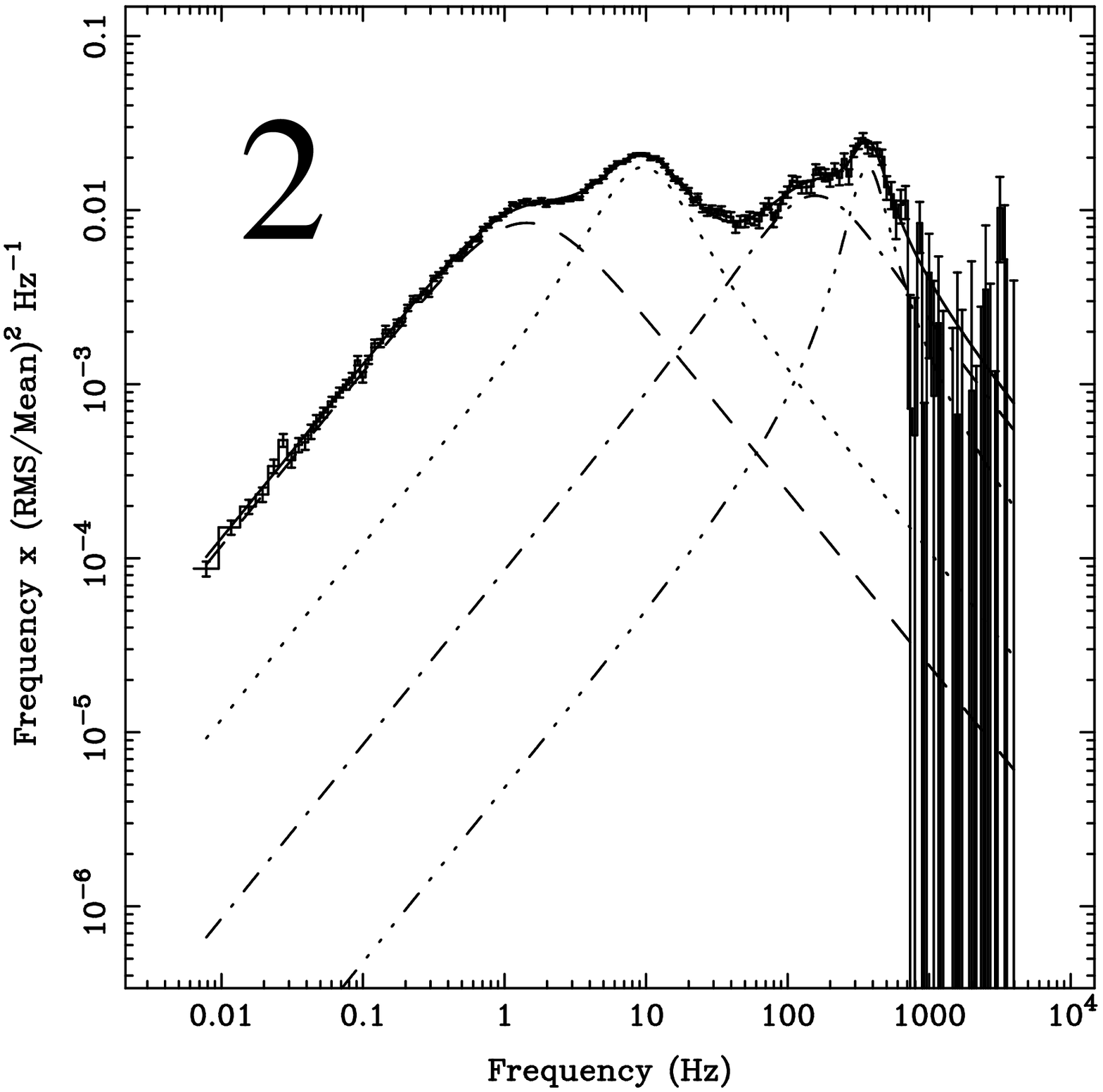} & \\
\plotone{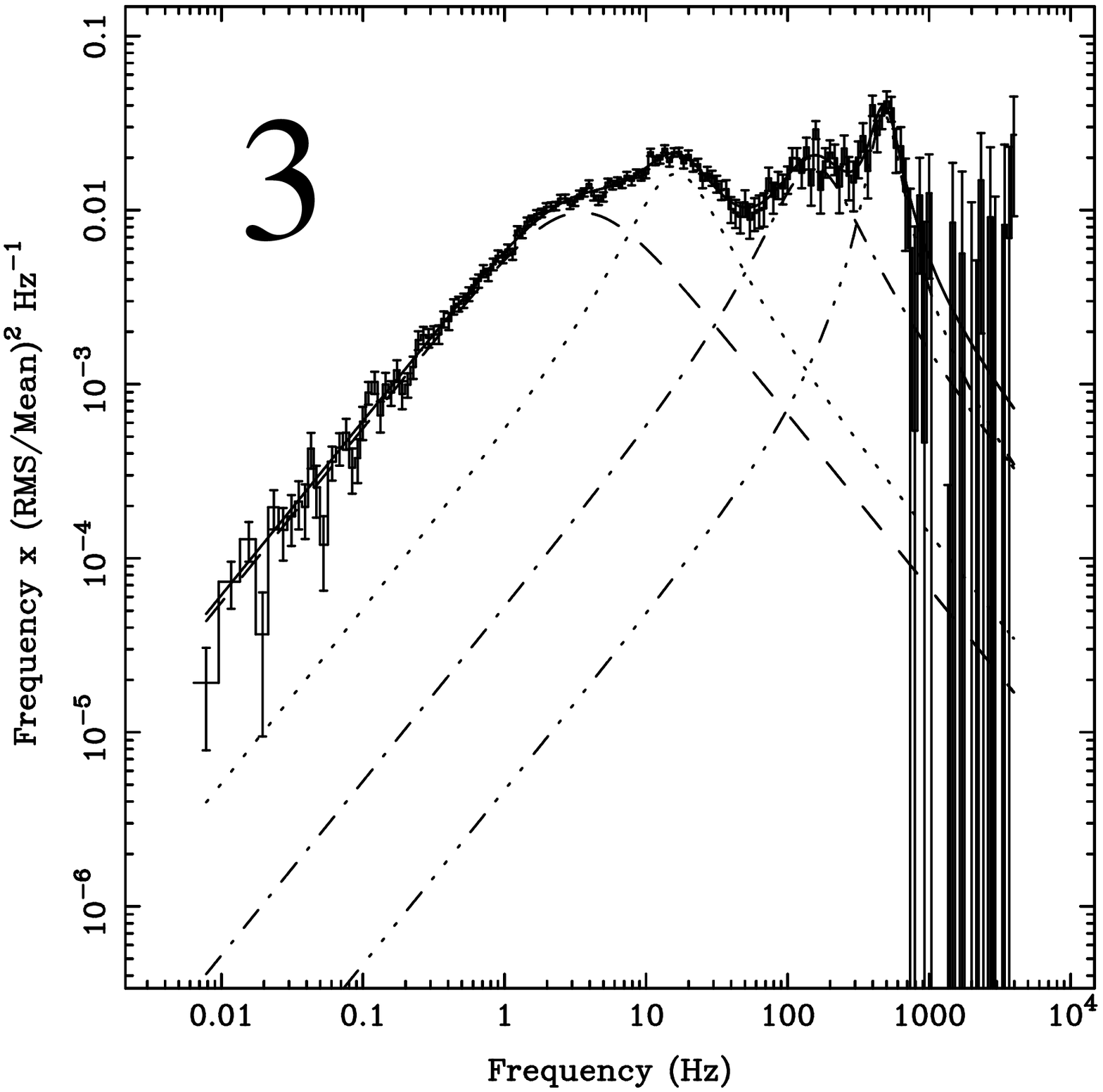} & 
\plotone{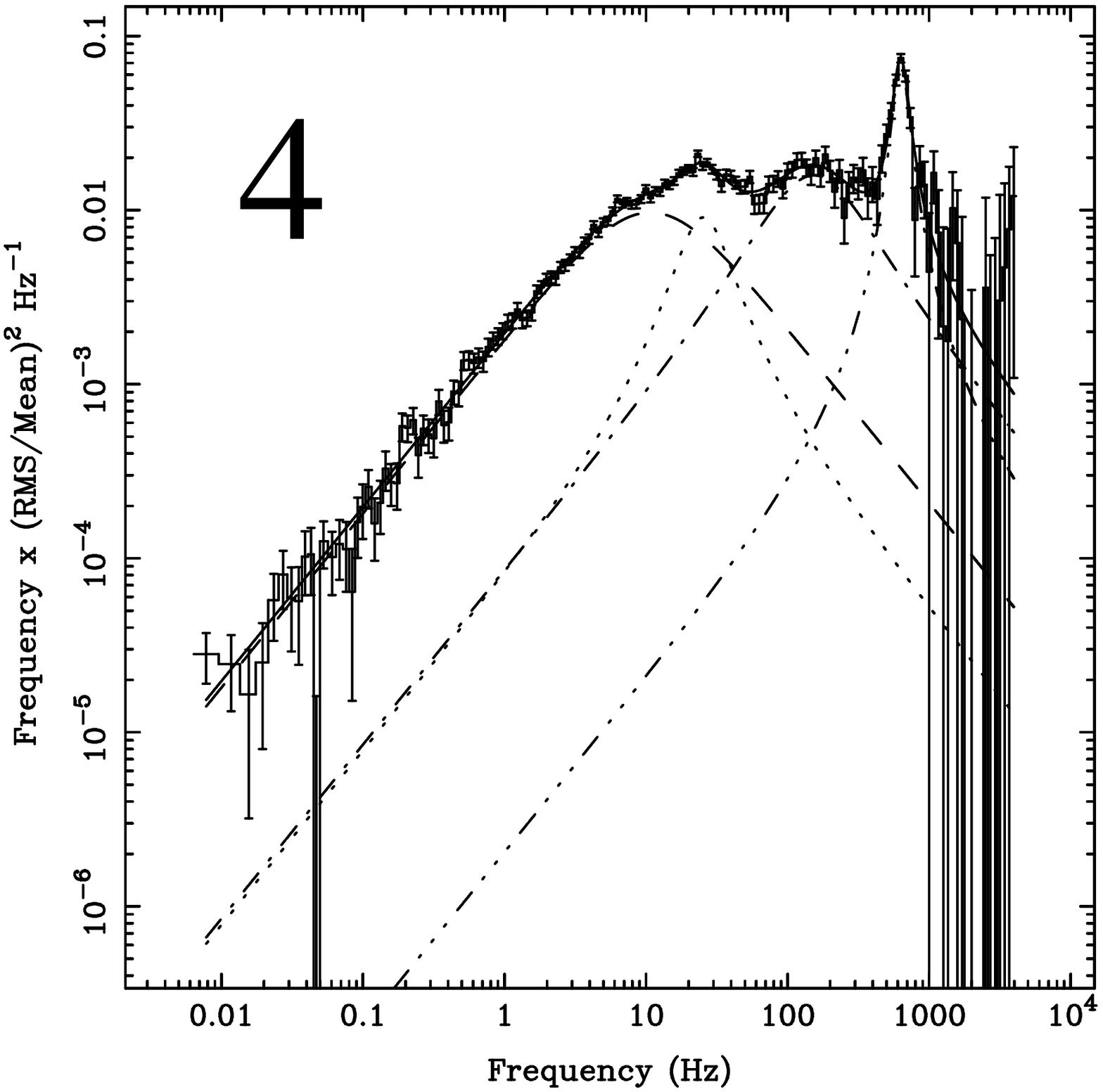} & \\
\end{tabular}
\caption{Power spectra and fit functions in the power spectral density 
times frequency representation (see \S 2) for 4U 0614+09. The different 
lines mark the individual Lorentzian components of the fit. The dashed 
lines mark both the BLN and the very low--frequency Lorentzian (\S 3.1), the dotted lines the 
low--frequency Lorentzian (\S 3.2), the dash--dotted line the hectohertz 
Lorentzian (\S 3.3) and the dash--dot--dot--dotted line both kilohertz 
QPOs (\S 3.4). In intervals 7 and 8 also a power--law is included to fit 
the VLFN. In interval 1 the identification of
the two highest frequency Lorentzians is unclear (see \S 3.4). 
Interval numbers are indicated.}
\label{fig.powspec_0614}
\end{figure}
\clearpage

\begin{figure}
\figurenum{2}
\epsscale{0.45}
\begin{tabular}{ccc}
\plotone{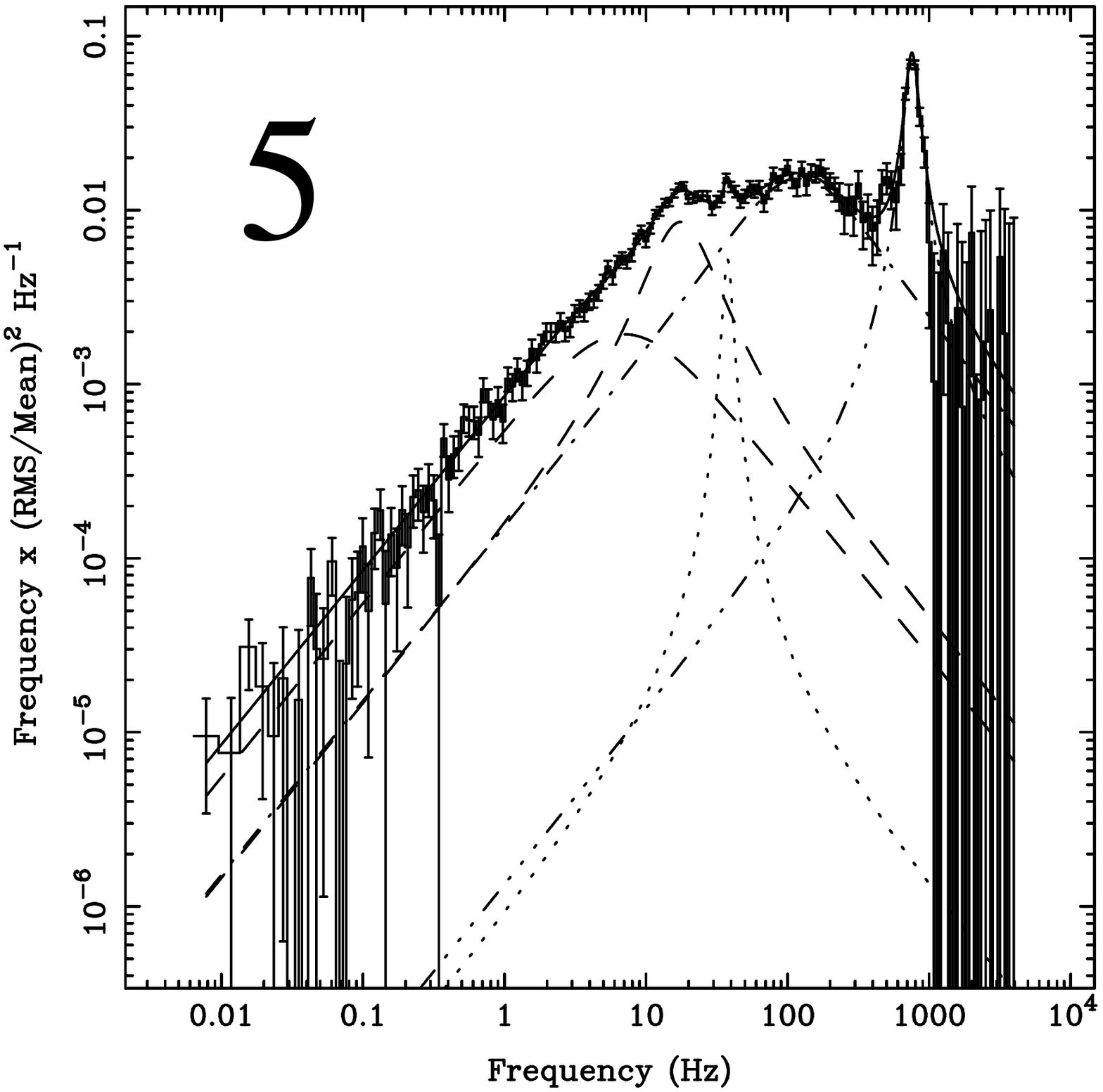} & 
\plotone{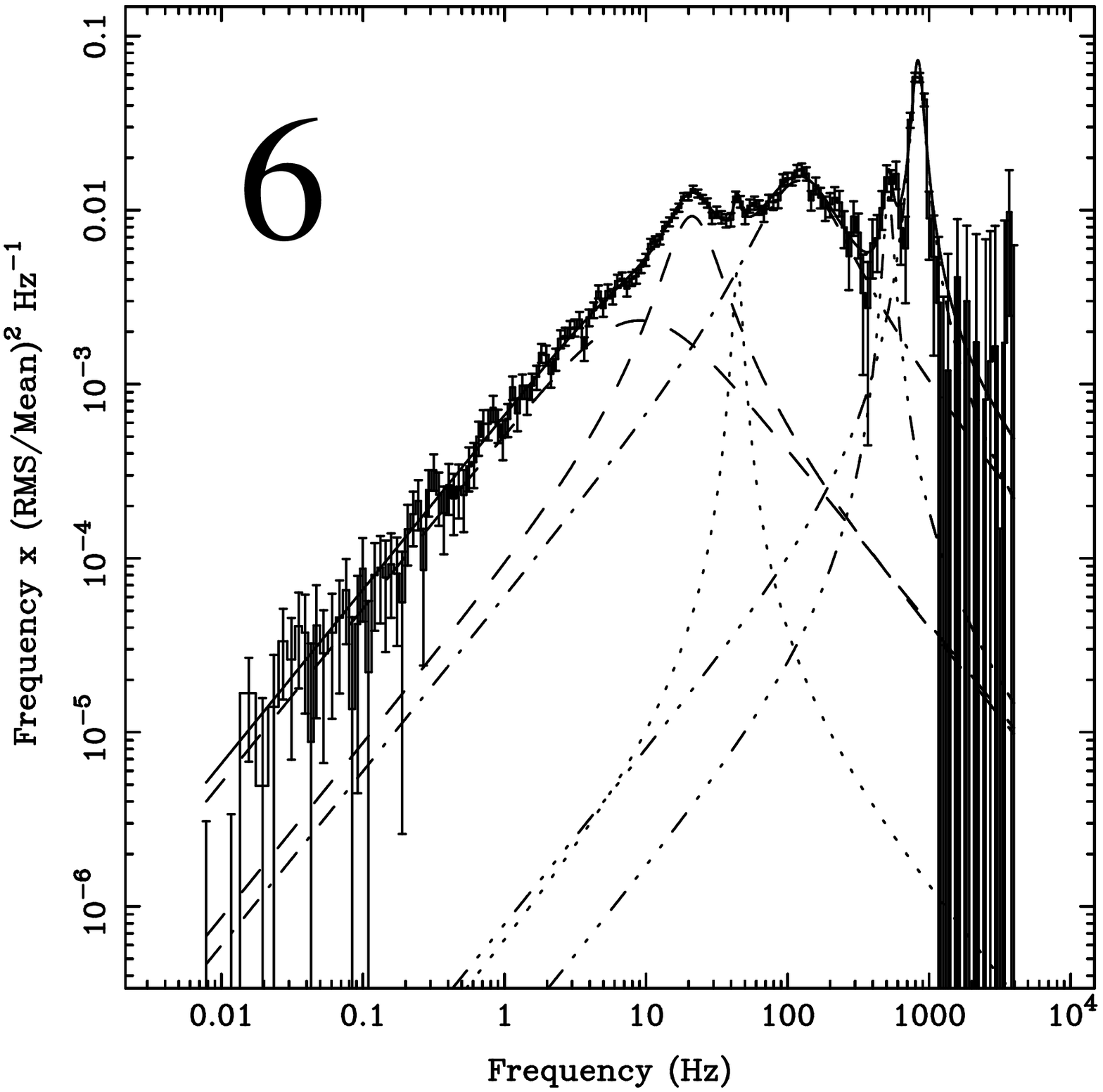} & \\
\plotone{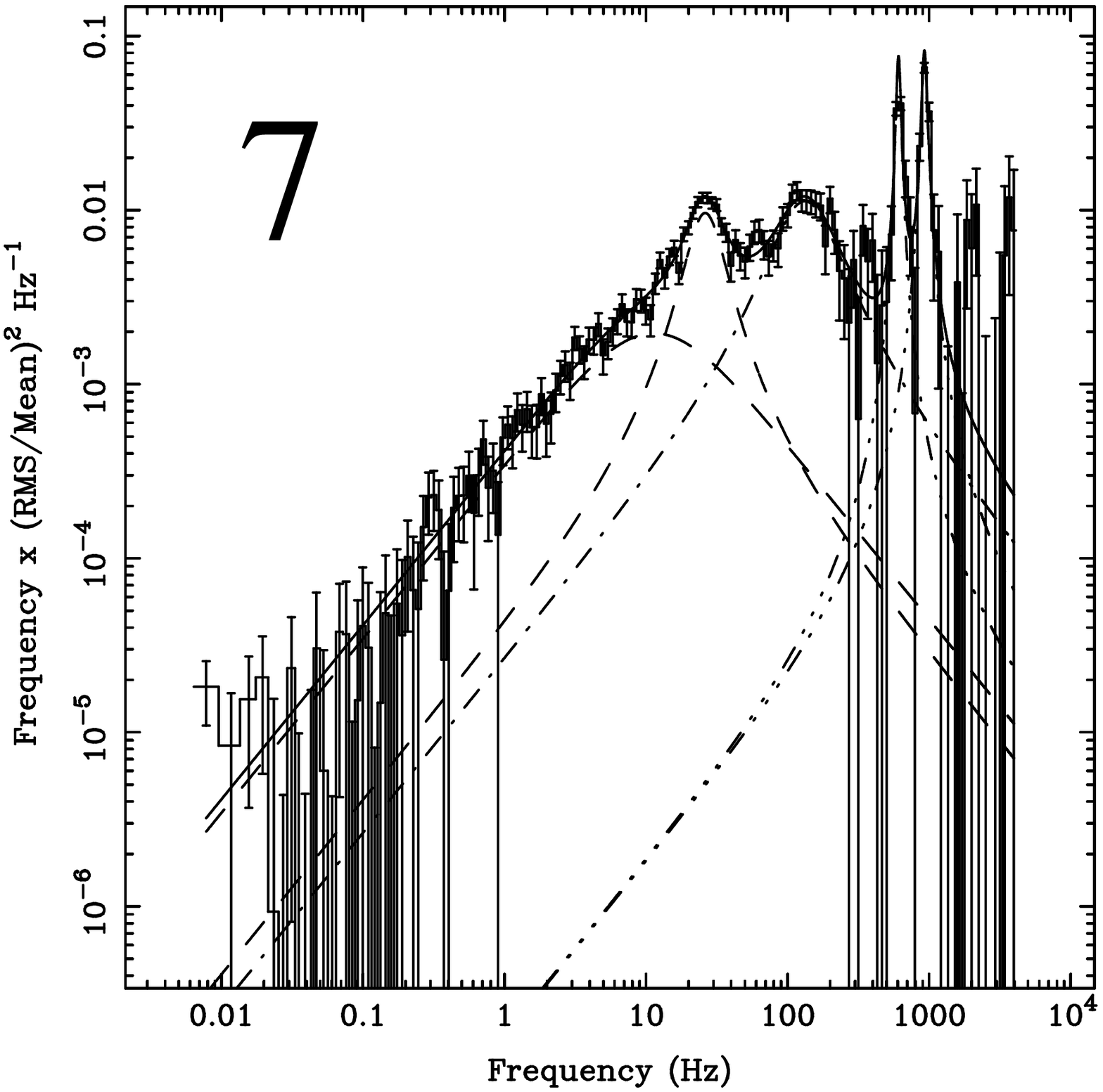} & 
\plotone{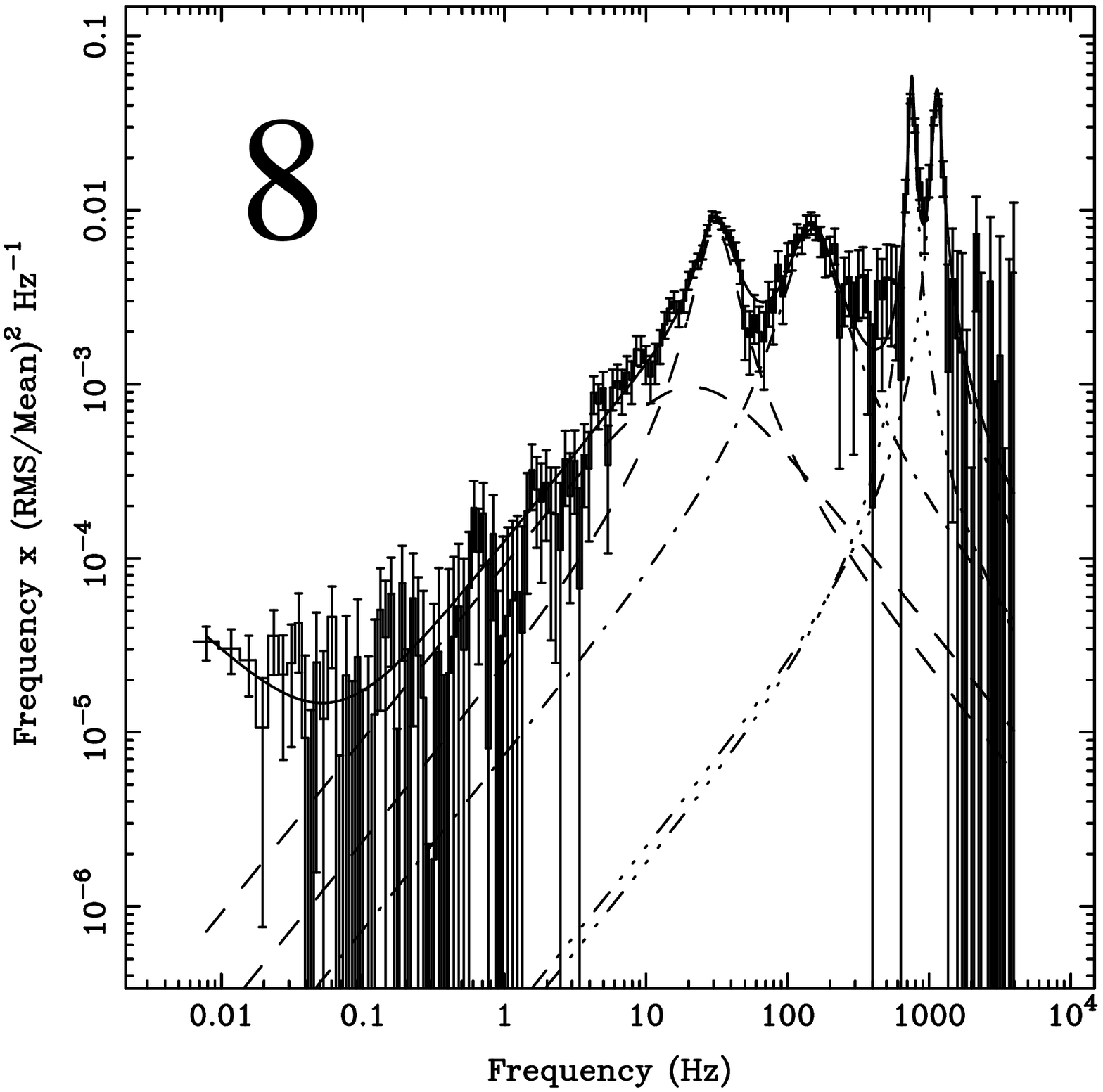} & \\
\end{tabular}
\caption{Continued}
\end{figure}
\clearpage

\begin{figure}
\figurenum{2}
\epsscale{0.45}
\begin{tabular}{ccc}
\plotone{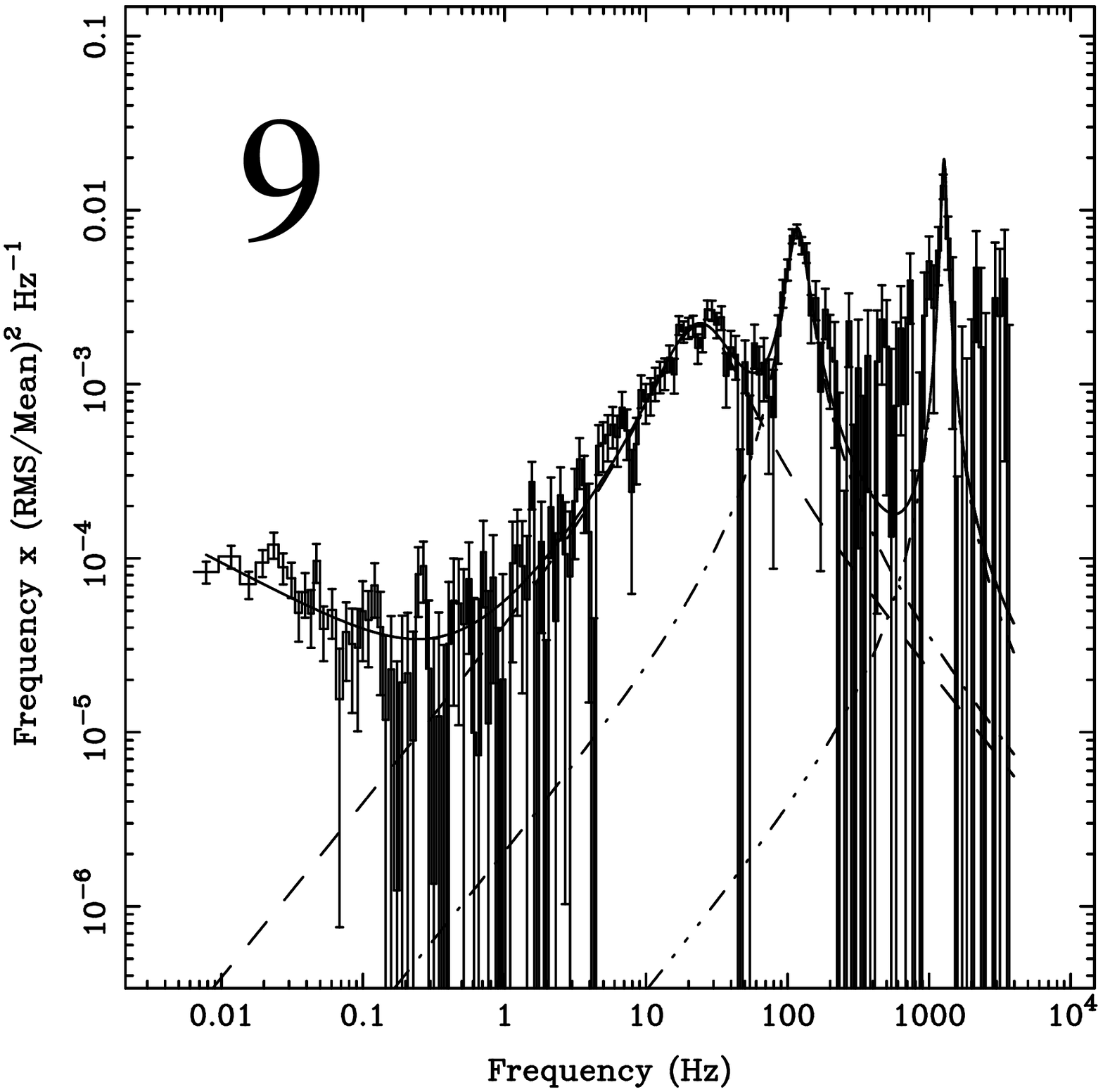} &  & \\
                    &  & \\
\end{tabular}
\caption{Continued}
\end{figure}
\clearpage

\begin{figure}
\figurenum{3}
\epsscale{0.8}
\plotone{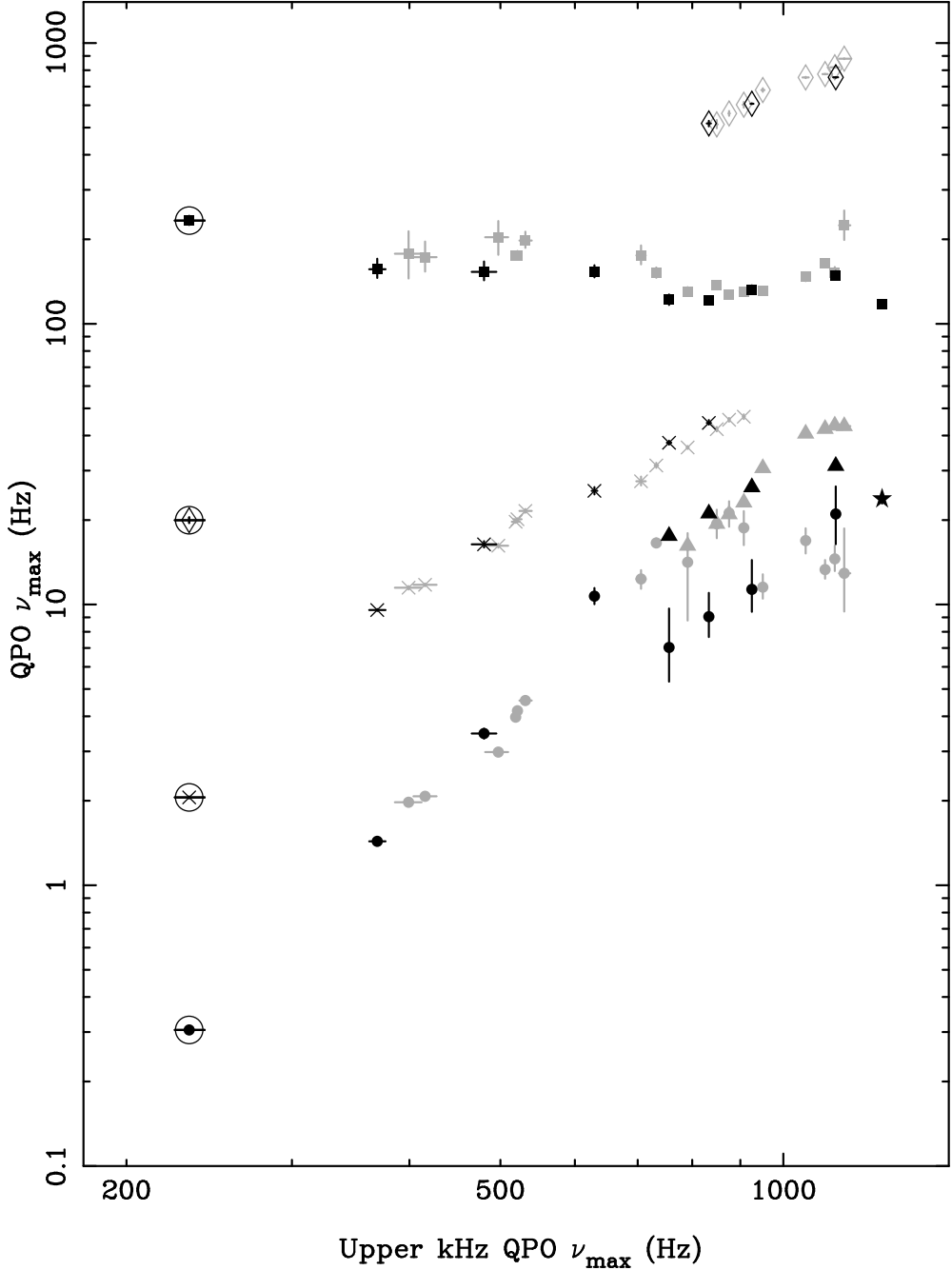}
\caption{Correlations between the characteristic frequencies 
($\equiv \nu_{\rm max}$) of the several Lorentzians used to 
fit the power spectra of 4U 1728--34 and 4U 0614+09 and the 
$\nu_{\rm max}$ of the Lorentzian identified as the upper 
kilohertz QPO. The points with the big circles on the very left 
are from interval 1 of 4U 0614+09, for which the fourth Lorentzian 
can be identified as either the upper kilohertz QPO or the 
hectohertz Lorentzian (see \S 3). The grey symbols are the 4U 1728--34 points, 
the black symbols the 4U 0614+09 points. The solid dots mark the 
BLN (zero--centered Lorentzian), the triangles the very low--frequency Lorentzian, the 
x--ses the low--frequency Lorentzian, the squares the hectohertz 
Lorentzian and the diamonds the lower kilohertz QPO. We use the parameters of 
this Lorentzian both for the upper kilohertz QPO and for the hectohertz 
Lorentzian. The star marks the one Lorentzian in interval 9 of 4U 0614+09 for 
which the identification is not clear (see \S 3.1). The other 
Lorentzians for which the identification is not clear (intervals 
18 and 19 of 4U 1728--34) are not in this plot as the upper 
kilohertz QPO is absent in those intervals.}
\label{fig.all_vs_upkilo}
\end{figure}
\clearpage

\begin{figure}
\figurenum{4}
\epsscale{0.9}
\plotone{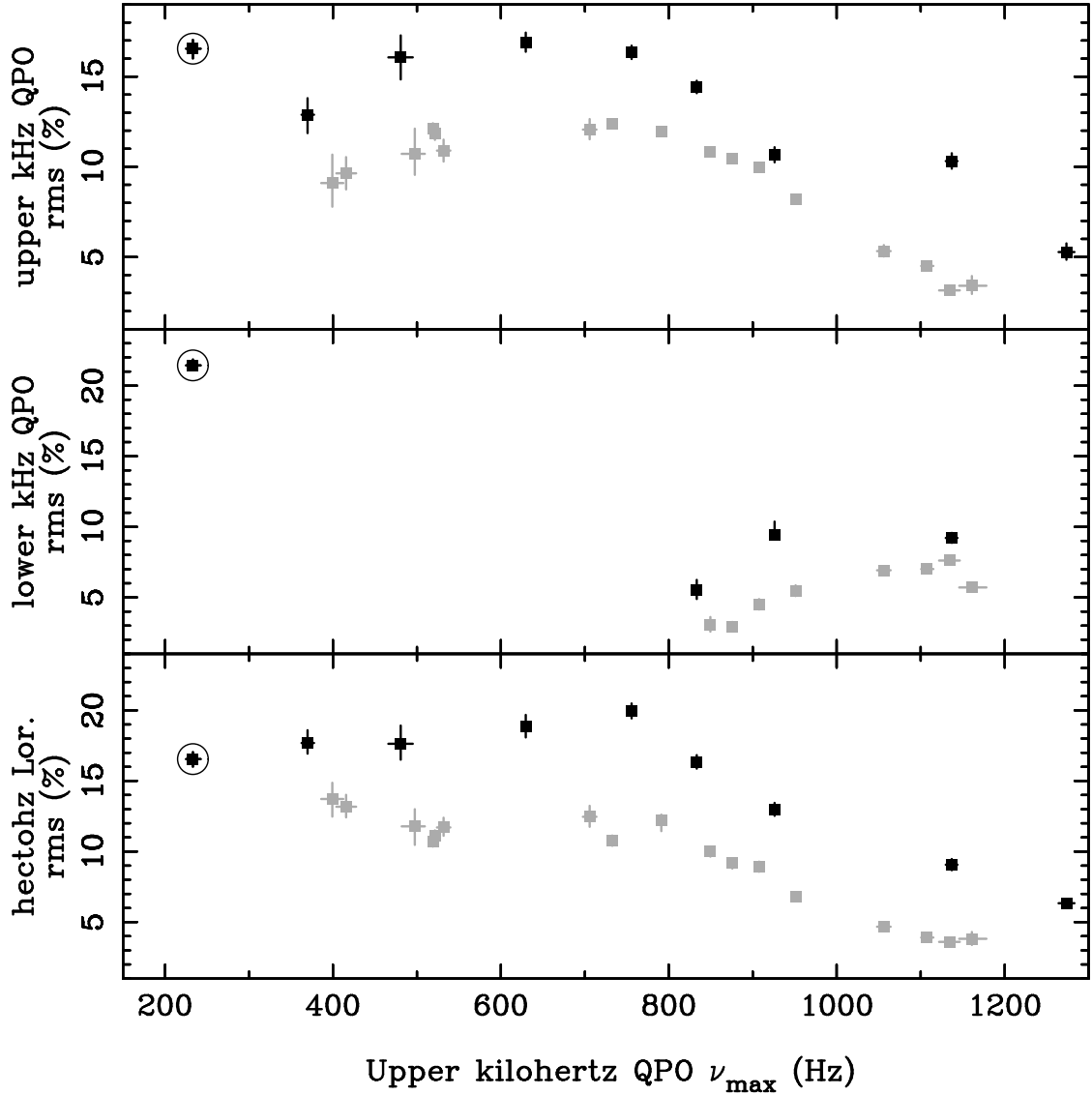}
\caption{The fractional rms (over the full PCA energy band)
of the hectohertz Lorentzian, the lower and upper kilohertz QPOs 
versus the $\nu_{\rm max}$ of the upper kilohertz QPO. The grey 
symbols are the 4U 1728--34 points, the black symbols the 4U 0614+09 points.
The circled points are from interval 1 of 4U 0614+09, for which 
the fourth Lorentzian can be identified as either the upper kilohertz 
QPO or the hectohertz Lorentzian (see \S 3). We use the parameters of 
this Lorentzian both for the upper kilohertz QPO and for the hectohertz 
Lorentzian.}
\label{fig.rms_vs_upkilo}
\end{figure}
\clearpage

\begin{figure}
\figurenum{5}
\epsscale{0.9}
\plotone{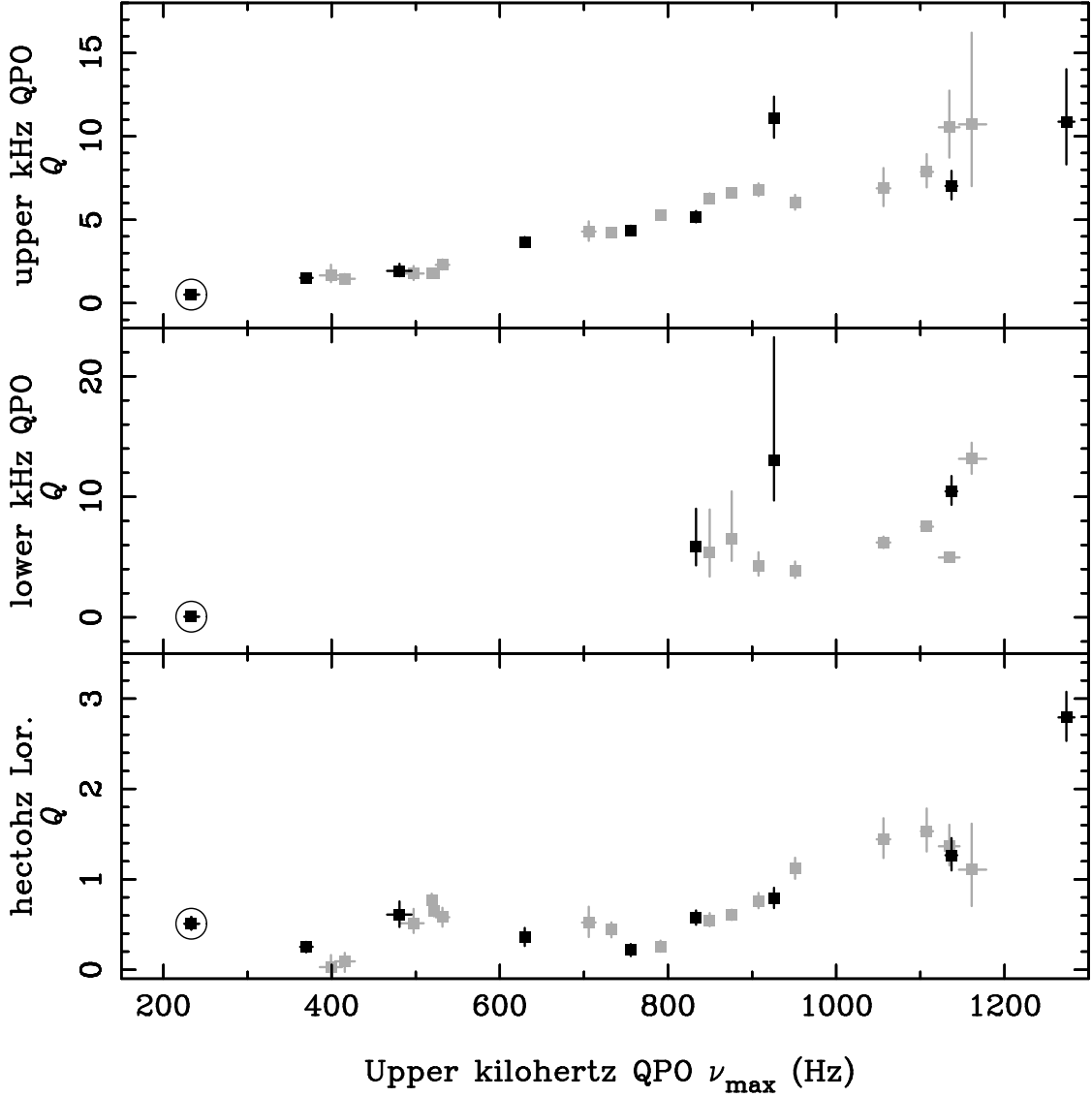}
\caption{The $Q$ values ($\equiv \nu_{\rm 0}/2\Delta$) of the 
hectohertz Lorentzian, the lower and upper kilohertz QPOs versus 
the $\nu_{\rm max}$ of the upper kilohertz QPO. The grey symbols 
are the 4U 1728--34 points, the black symbols the 4U 0614+09 points.
The circled points are from interval 1 of 4U 0614+09, for which 
the fourth Lorentzian can be identified as either the upper kilohertz 
QPO or the hectohertz Lorentzian (see \S 3). We use the parameters of 
this Lorentzian both for the upper kilohertz QPO and for the hectohertz 
Lorentzian.}
\label{fig.q_vs_upkilo}
\end{figure}
\clearpage

\begin{figure}
\figurenum{6}
\epsscale{0.9}
\plotone{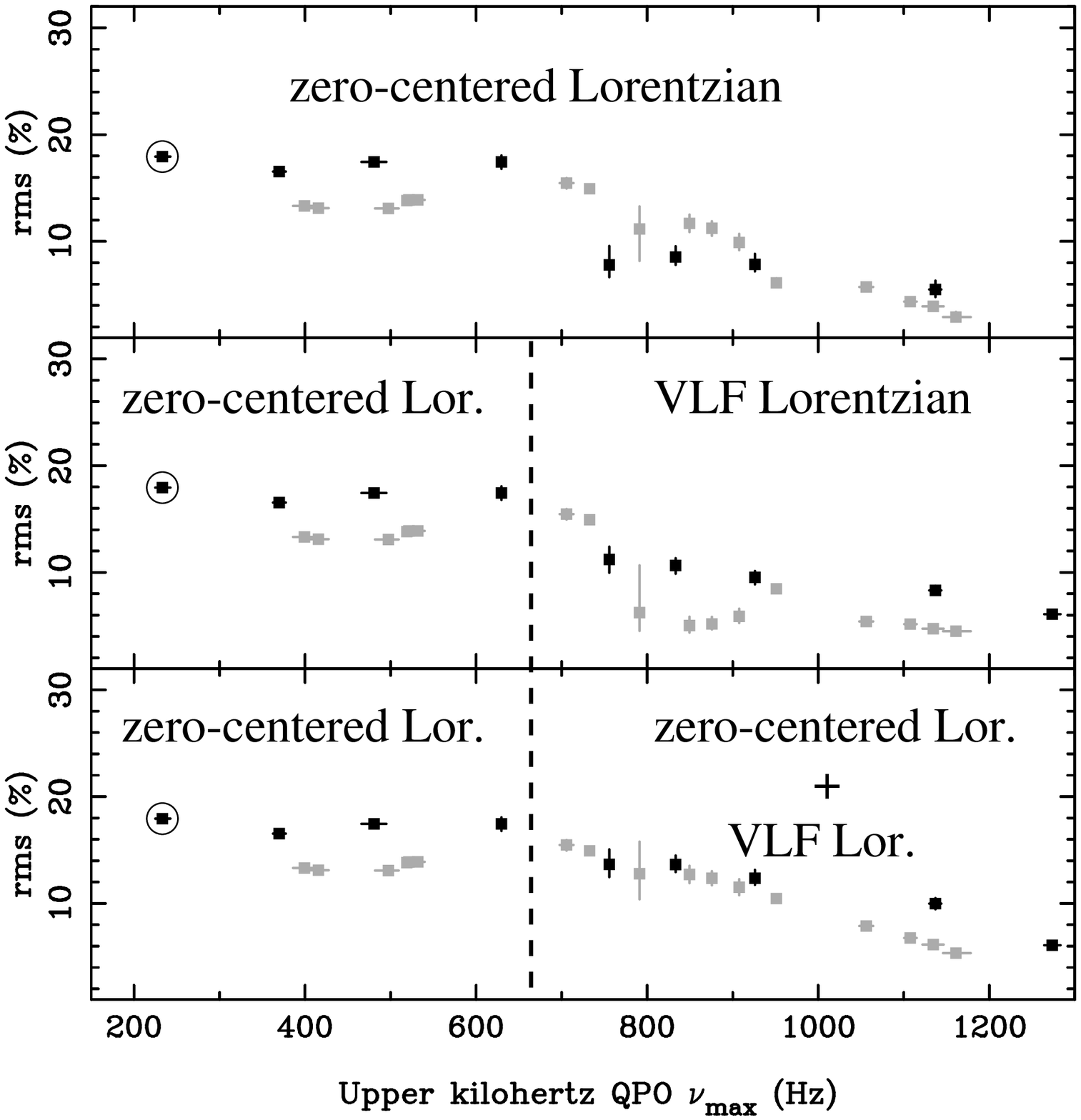}
\caption{The fractional rms of the BLN versus $\nu_{\rm upper kHz}$. 
In the top panel we plot the rms of the zero--centered Lorentzian vs. $\nu_{\rm upper kHz}$. 
In the top panel we plot the fractional rms of the zero--centered 
Lorentzian. In the middle panel we plot the fractional rms of the very low--frequency Lorentzian when 
it is present, and the fractional rms of the zero--centered Lorentzian when it is not. In 
the bottom panel finally we plot the fractional rms of the zero--centered 
Lorentzian and the very low--frequency Lorentzian together. The grey symbols mark the 4U 1728--34 
points, the black symbols the 4U 0614+09 points. 
The circled points are from interval 1 of 4U 0614+09, for which 
the fourth Lorentzian can be identified as either the upper kilohertz 
QPO or the hectohertz Lorentzian (see \S 3). We use the parameters of 
this Lorentzian both for the upper kilohertz QPO and for the hectohertz 
Lorentzian.}
\label{fig.upkilo_vs_bln}
\end{figure}
\clearpage

\begin{figure}
\figurenum{7}
\epsscale{1.0}
\plotone{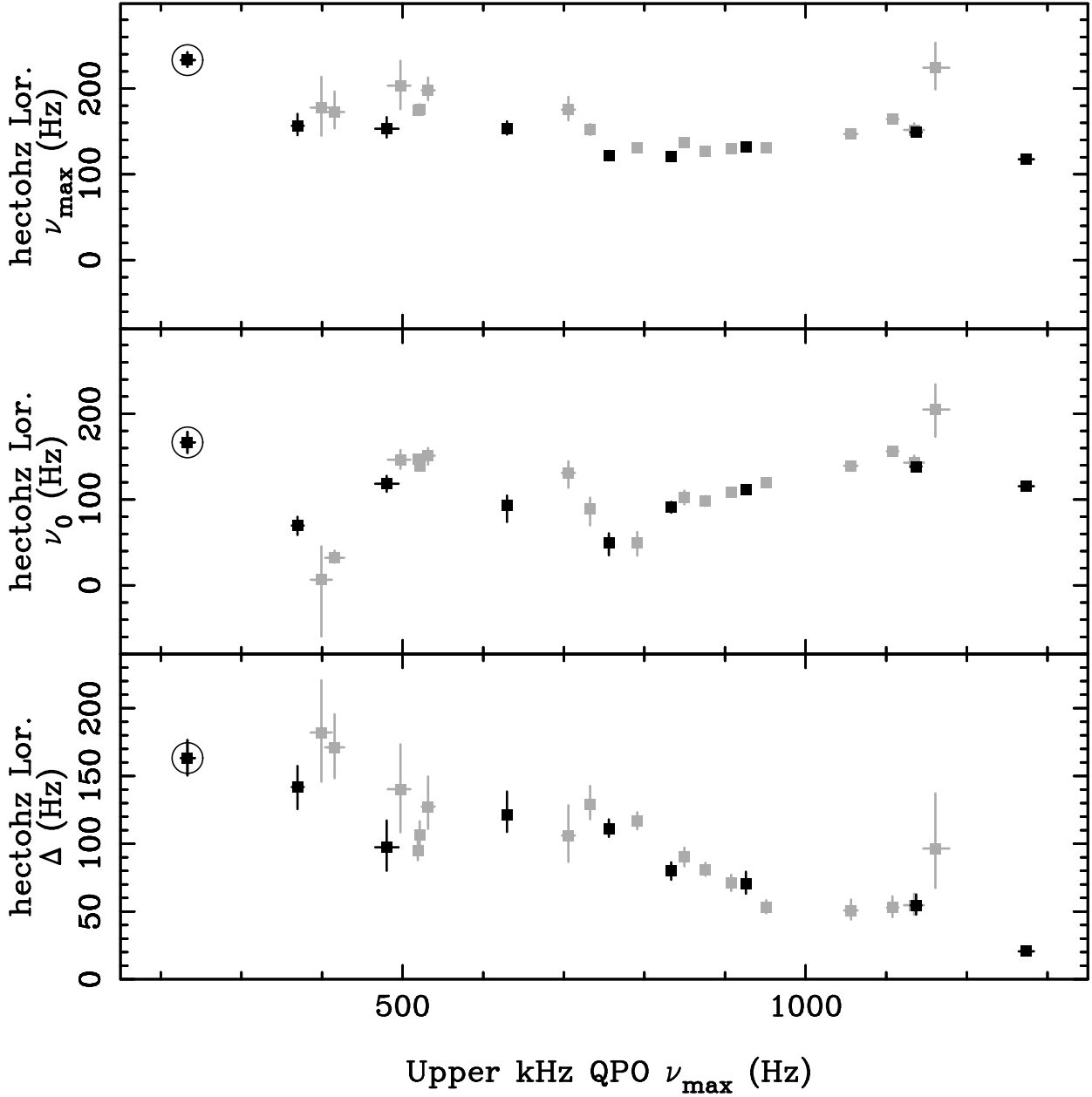}
\caption{$\nu_{\rm max}$, $\nu_{\rm 0}$ and $\Delta$ of the hectohertz 
Lorentzian versus the $\nu_{\rm max}$ of the upper kilohertz QPO. 
The grey symbols mark the 4U 1728--34 points, the black symbols the 4U 
0614+09 points. The characteristic frequency of the hectohertz Lorentzian 
is either dominated by $\nu_0$, $\Delta$ or both and therefore the use 
of $\nu_{\rm max}$ as the characteristic frequency of the hectohertz 
Lorentzian most clearly shows the near constant frequency of this feature. 
For comparison the top two panels are plotted on the same scale.
The circled points are from interval 1 of 4U 0614+09, for which 
the fourth Lorentzian can be identified as either the upper kilohertz 
QPO or the hectohertz Lorentzian (see \S 3). We use the parameters of 
this Lorentzian both for the upper kilohertz QPO and for the hectohertz 
Lorentzian.}
\label{fig.hecto}
\end{figure}
\clearpage

\begin{figure}
\figurenum{8}
\epsscale{0.77}
\plotone{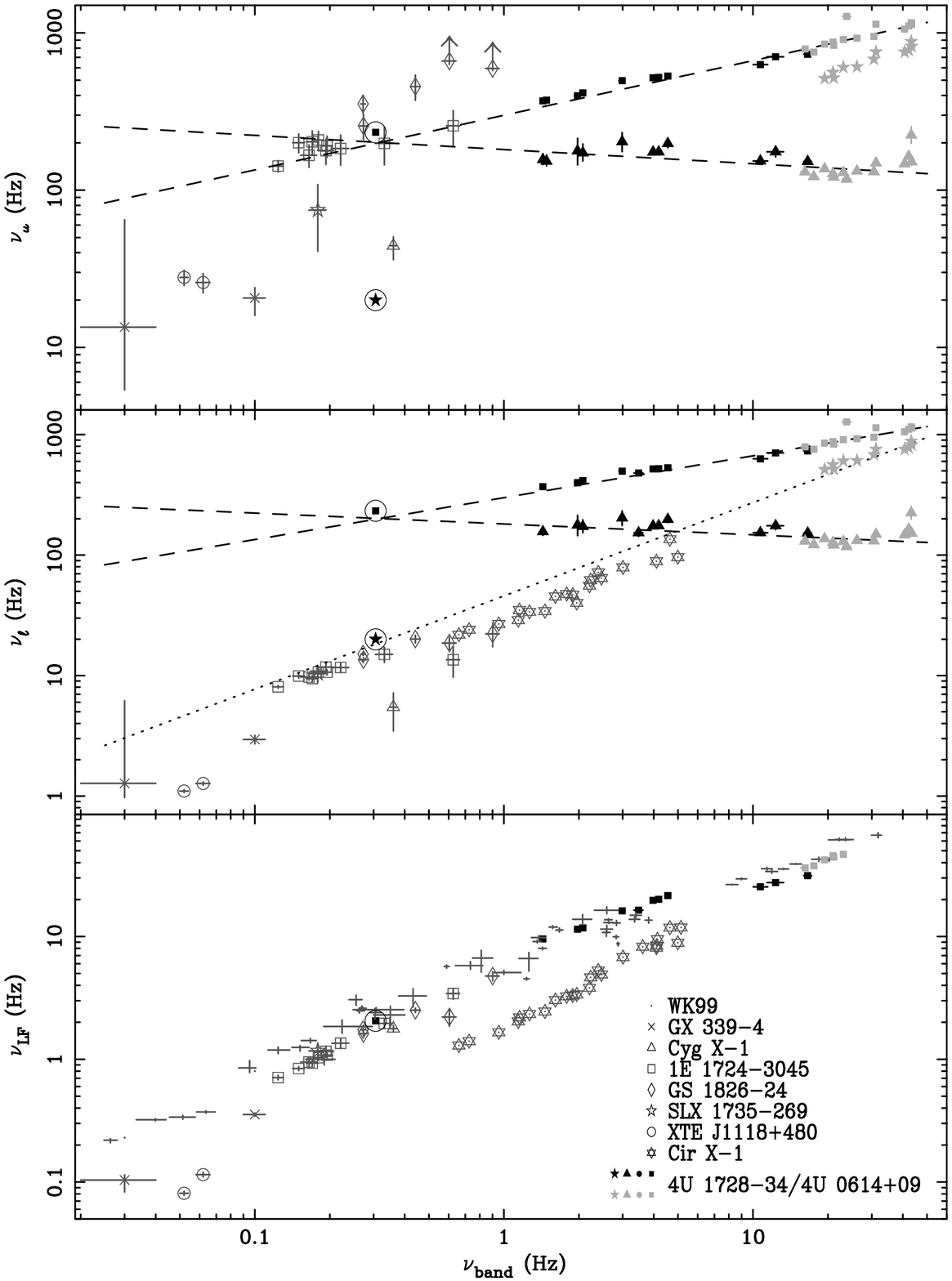}
\caption{Comparison of the characteristic frequencies of the 
multi--Lorentzian fit to 4U 0614+09 and 4U 1728--34 with several 
other LMXBs. The open grey symbols represent the results from Shirey 
(1998), Nowak (2000) and BPK01, the filled symbols our results 
of 4U 1728--34 and 4U 0614+09. The filled black points mark the points 
where we used $\nu_{\rm BLN}$ for $\nu_{\rm band}$, the filled grey points 
those where we used $\nu_{\rm VLF}$ (see \S 4).
The symbols representing the different 
sources are indicated in the plot. For 4U 1728--34 and 4U 0614+09 
the squares mark the upper kilohertz QPOs, the stars the lower 
kilohertz QPO, the triangles the hectohertz Lorentzian and the 
circles the low--frequency Lorentzian. For clarity the figure is 
split up in three panels. In the bottom panel we plot the 
characteristic frequency of the first Lorentzian versus the 
characteristic frequency of the band--limited noise (the WK99 
relation). In the middle and top panels we plot the characteristic 
frequencies of the hectohertz Lorentzian and both kilohertz QPOs 
versus the characteristic frequency of the band--limited noise for 
4U 1728--34 and 4U 0614+09. The dashed lines indicate extrapolated 
power--law fits (see \S 4) to the 4U 1728--34 and 4U 0614+09 points.
In the midle panel we compare with the characteristic frequency of 
the second Lorentzian ($\nu_\ell$ in PBK01).
The dotted line in the middle panel indicates a power--law fit 
to the lower kilohertz QPO points of 4U 1728--34 and 4U 0614+09 and 
the second Lorentzian points of the low--luminosty bursters (see \S 5.3). 
In the top panel we compare
with the third Lorentzian ($\nu_u$ in PBK01) of Nowak (2000) and 
BPK01. The arrows in the top panel indicate lower limits.
The circled points are from interval 1 of 4U 0614+09, for which 
the fourth Lorentzian can be identified as either the upper kilohertz 
QPO or the hectohertz Lorentzian (see \S 3).}
\label{fig.comp}
\end{figure}
\clearpage

\begin{figure}
\figurenum{9}
\epsscale{0.72}
\plotone{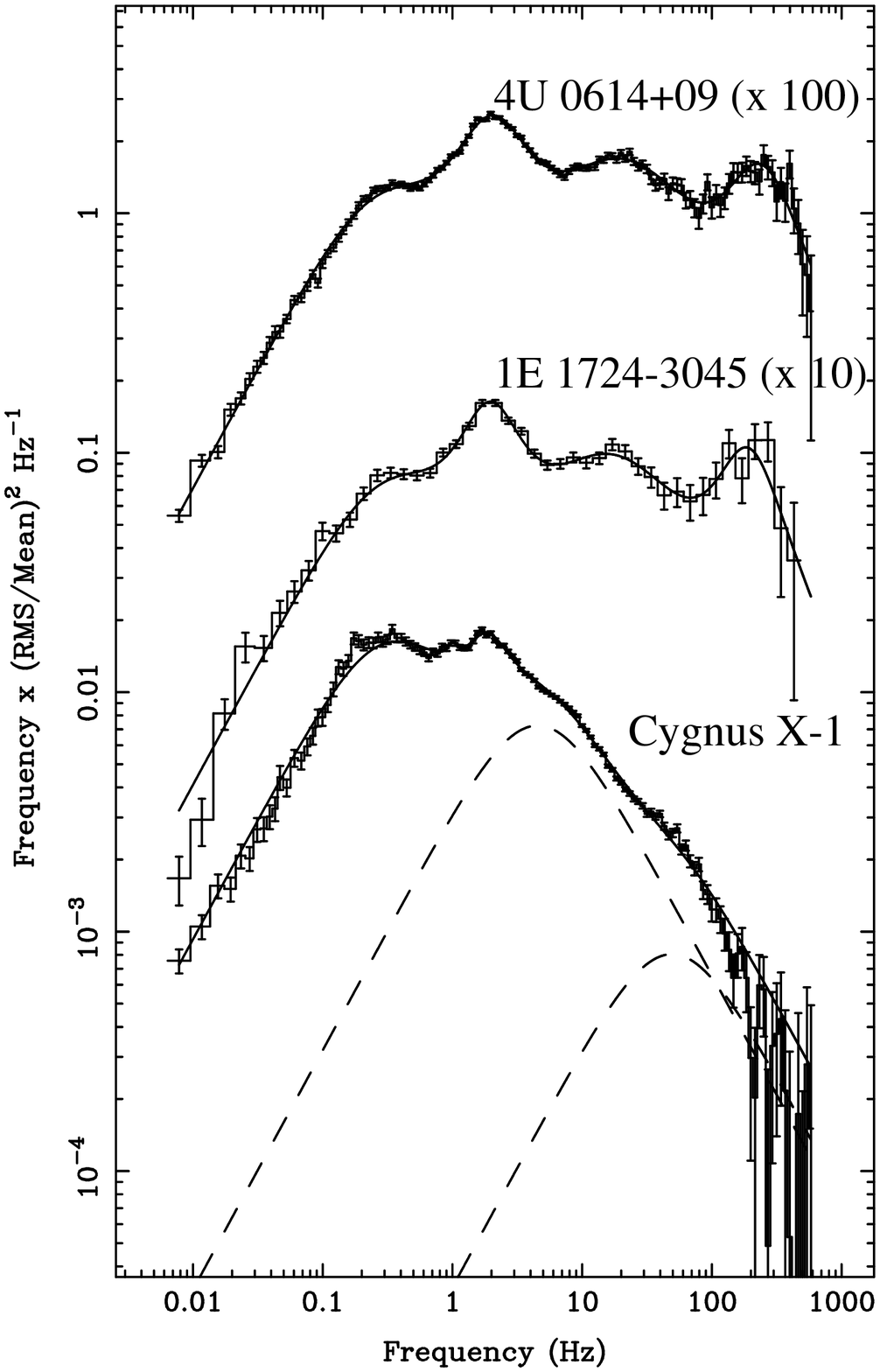}
\caption{Power spectra and fit functions of 4U 0614+09, 1E 1724--3045 and Cyg X--1 all with a 
$\nu_{\rm band}$ of about 0.3 Hz. For clarity the power spectra of 4U 0614+09 and 1E 1724--3045 are 
multiplied with a constant as indicated in the plot. The dashed lines in the power spectra of Cyg X--1 
show the individual two high frequency Lorentzians.}
\label{fig.cygx1_1E1724_4U0614}
\end{figure}
\clearpage

\end{document}